\documentclass[12pt]{article}

\usepackage{amsmath,natbib}
\usepackage{amssymb,color}
\usepackage{appendix}
\usepackage{amsthm}
\usepackage{enumitem}
\usepackage{hyperref}
\usepackage{mathtools} 
\mathtoolsset{showonlyrefs=true}  
\usepackage{bm}
\usepackage{setspace}

\usepackage[margin=0.95in]{geometry}

\numberwithin{equation}{section}

\theoremstyle{plain}

\newtheorem{definition}{Definition}[section]

\newtheorem{proposition}{Proposition}[section]

\begin{document}

\author{Tim Leung\thanks{Department of Applied Mathematics, Computational Finance and Risk Management Program, University of Washington, Seattle WA 98195. E-mail:
\mbox{timleung@uw.edu}.  Corresponding author.}  \and Hyungbin Park\thanks{Department of Mathematical Sciences, Worcester Polytechnic Institute, Worcester, MA 01609, USA.    E-mail: hpark@wpi.edu.}}
\title{Long-Term Growth Rate of Expected Utility\\ for Leveraged ETFs: Martingale Extraction Approach}
\date{\today}   

\maketitle

\begin{abstract}
This paper studies the  long-term growth rate  of   expected utility  from holding a leveraged exchanged-traded fund (LETF), which is a constant proportion portfolio of the reference asset.  Working with the power utility function, we develop an analytical approach that employs martingale extraction and   involves   finding the eigenpair associated with the infinitesimal generator of a Markovian time-homogeneous  diffusion.  We derive    explicitly the long-term growth rates  under a number of models for the reference asset,  including the geometric Brownian motion model,  GARCH model, inverse GARCH model, extended CIR model, 3/2 model, quadratic model,  as well as  the Heston and $3/2$ stochastic  volatility models. We also investigate the impact of stochastic interest rate such as the Vasicek model and the inverse GARCH short rate model. We  determine  the optimal leverage ratio for the long-term investor and examine the effects of model parameters. \end{abstract}

\newpage
\tableofcontents
\newpage

\section{Introduction} 
Exchange-traded funds (ETFs) are popular financial products designed  to track the value of a reference asset or index. With over \$2 trillion of  assets under management, ETFs  are traded on major exchanges  like  stocks, even if the reference itself may not be traded. Within the growing ETF market, leveraged ETFs (LETFs)  are created to generate a  constant  multiple $\beta$, called leverage ratio, of the daily returns of a reference index.  For example, the ProShares Ultra S\&P 500 (SSO)  offers to  generate twice ($\beta =2$) the daily returns of  the S\&P 500 index.   In the LETF market, the most common leverage ratios are $\beta\in \{1,2,3\}$ and $\beta\in\{-1,-2,-3\}$.  In particular, investors can take a bearish position on the reference    by taking a long position in an LETF with $\beta< 0$ without   the need of borrowing shares or  a margin account.   For many speculative investors, LETFs are  highly accessible and liquid instruments that give a leveraged exposure, and particularly attractive  during periods of large momentum.

For LETF holders and potential investors, it is crucial importance  to understand the price dynamics and the impacts of leverage ratio on the risk and return of each LETF. A number of market observations suggest that LETFs suffer from the volatility decay effect, which reflects the  value erosion   proportional to the realized variance  of the reference index. Recent studies, including  \cite{AZLETF}, \cite{ChengLETF}, \cite{LeungWard},  and \cite{Leung2016},  present  discrete-time and continuous-time stochastic frameworks to illustrate the the path dependence of LETFs on the reference, including the  volatility decay effect. In fact, SEC   issued in 2009 an alert announcement regarding the   riskiness of LETFs, especially when holding them long-term.\footnote{See the SEC alert on http://www.sec.gov/investor/pubs/leveragedetfs-alert.htm.}   \cite{LeungSantoli} derived  the  admissible holding horizons  for LETFs with respect to  different risk measures.   This motivates us to analyze the long-term growth rate of expected utility of holding an LETF, and examine the dependence on the leverage ratio and dynamics of the reference.

 In this paper, we investigate   the long-term growth rate of  the expected utility from holding a LETF. Specifically, we consider different   stochastic models for the      LETF price, denoted by $L_t$, and analyze the long-term growth rate represented by the limit:
\begin{align}\lim_{t\rightarrow\infty}\frac{1}{t}\log\mathbb{E}[u(L_t)]\;,\label{problem1}\end{align}
where $u(\cdot)$ is the investor's utility   of the power form: $u(w)=w^\alpha$ with $0<\alpha\leq 1$. As such, the   coefficient of relative risk aversion is given by $\varrho:=- w u''(w) / u'(w) = 1-\alpha$, $\forall w>0$. When $\alpha=1$, corresponding to zero relative risk aversion,  the limit is  the  long-term  growth rate of \emph{expected return} of  the LETF. Hence, analyzing \eqref{problem1} allows us to understand the long-term growth rates of expected utility and expected return useful for   risk-averse  and risk-neutral investors, respectively. 

One of main contributions of this paper is to present a novel approach to determine the above limit  analytically. For this purpose, we  employ the method of \emph{martingale extraction}, through which 
the problem of finding the  long-term growth  
is transformed into the eigenpair (eigenvalue and eigenfunction) problem of a second-order differential operator that is associated with the infinitesimal generator of the  reference   process.  

Our results allow us to determine the  optimal  leverage ratio for the long-term risk-averse  investor. For the  $\beta$-LETF with price denoted by $L_t\equiv L_t{(\beta)}$, we   find the optimal leverage ratio  $\beta^*$ that maximizes the long-term growth rate, that is,
\begin{align}\label{maxbeta}\beta^* = \arg\max_{\beta\in\mathbb{R}}\,\lim_{t\rightarrow\infty}\frac{1}{t}\log\mathbb{E}[u(L_t{(\beta)})]\,. \end{align}
Furthermore, we examine through our explicit expressions the combined effects  of risk aversion and model parameters  on the optimal choice of leverage.

There are a number of related studies on the  long-term growth rate of expected utility.  The  seminal work by \cite{Fleming1999} investigated the 
optimal growth rate of expected utility of wealth. The utility was of hyperbolic absolute risk aversion (HARA) type, and  dynamic programming scheme was developed for  different HARA parameters and  policy constraints.  \cite{akian1999} studied the optimal investment strategies with transaction costs with the    objective   to maximize
the long-term average growth rate  under logarithmic utility.  In another related study, \cite{Zhu2014} also examined  the  long-term growth rate  of expected power utility from a nonleveraged portfolio with a fixed fraction of wealth in the single risky asset, and  derived explicit limits under some  models. In comparison, we study \textit{leveraged}  portfolios under additional single-factor and multi-factor diffusion models. 

\cite{christensen2012} considered  the  growth rate maximization  problem based
on impulse control strategies with limited number of trades per unit time and proportional transaction costs.  
\cite{guasoni2015limits} analyzed the optimal strategy to maximize the long-term return given average volatility under  the    Black-Scholes model with proportional costs. 
\cite{hata2006solving} studied  a long-term optimal investment problem   with an objective of maximizing the probability that the portfolio value would exceeed  a given   level  in  a market  with  Cox-Ingersoll-Ross  interest rate. 
Applying the theory of  large deviation, \cite{Pham2003} derived the  optimal long-term investment strategy under the  CARA  utility, and  \cite{pham2015} examined the long-term asymptotics for optimal portfolios that involved   maximizing the probability for a portfolio to  outperform a target growth rate.

The martingale extraction method is a relatively new  analytical technique that has been used to investigate a number of   financial and economic problems. Among our main references, \cite{Hansen2009} and \cite{Hansen2012}
developed the  martingale extraction method to study the  long-term risk in continuoue-time Markovian markets.  \cite{BorovickaHansenHendricksScheinkman2011} utilized  the martingale extraction method to examine  the  shock exposure in terms of shock elasticity that  measures the impact of  shock. In these studies, the authors  decompose   a pricing operator into   three components: an exponential term, a martingale and a transient term, each of which    carries a financial interpretation depending on the context of problem. \cite{Park2016} studied    sensitivities   of long-term cash flows with respect to perturbations of underlying processes by using the martingale extraction method. 
\cite{QinLinetsky2015} further analyzed the Hansen-Scheinkman factorization (martingale extraction)   for positive eigenfunctions of Markovian pricing operators. Our contribution on this front is to be the first to apply the martingale extraction technique to compute explicitly  the long-term growth rate of expected utility.

The rest of this paper proceeds as follows. In Section \ref{sec:LETF_dynamics}, we discuss our   martingale extraction approach for LETFs. In Section \ref{sec:Univari}, we solve the long-term growth rate problem when  the reference   price  follows a one-dimensional Markov diffusion.  Sections \ref{sec:SV}, \ref{sec:interest_rate}  and \ref{sec:Q} are dedicated to, respectively, stochastic volatility models, interest rate models, and quadratic models. We compute the long-term growth rates and determine the optimal leverage ratios.   Section \ref{sec:conclusion} summarizes this paper.

\section{Martingale extraction approach for LETFs}
\label{sec:LETF_dynamics}

Let $(\Omega,\mathcal{F},\mathbb{P})$ be a  probability space where $\mathbb{P}$ is the
subjective probability measure. Denote by  $\mathbb{F} \equiv (\mathcal{F}_t)_{t\ge 0}$ the filtration    generated by  a
$d$-dimensional standard Brownian motion $B$.  Consider a     reference index, such as the S\&P500 index, whose    price process  $X$ is a one-dimensional positive time-homogeneous Markov diffusion process satisfying  \footnote{Throughout, we  use the dot notation $\,\cdot\,$ for the multiplication of column vectors, and omit the dot for the matrix multiplication.} 
\begin{align}\label{dXt1}
\frac{dX_t}{X_t}=\mu_t\,dt+\sigma_t\cdot dB_t\,,  \qquad t\ge 0\;,
\end{align}
where  the   drift process  $\mu_t$ and     vector volatility process  $\sigma_t$  are both   $\mathbb{F}$-adapted.
At this point, we do not specify a
parametric stochastic drift or volatility model, though many well-known models, such as the
Heston model as well as other stochastic or local volatility models, also fit within the above framework. In addition,  the  risk-free rate process is denoted by   $(r_t)_{t\ge 0}$ which may be constant or stochastic depending on the model.

\subsection{LETF price dynamics}
A leveraged ETF is a constant proportion portfolio in the reference $X$.   A long-LETF based  on $X$ has  a leverage ratio $\beta\geq 1$. At any time $t,$ the cash amount of $\beta L_t$ ($\beta$ times the fund value) is
invested in $X$ and  the amount $(\beta-1)L_t$ is borrowed at the risk-free rate $r_t$. Strictly speaking, for $\beta\in[0,1)$, the fund is not leveraged since only a fraction of the fund value is invested in the risk asset, and   no money is borrowed. For a short-LETF   with ratio $\beta <0$,  a short position of the  amount   $|\beta|L_t$ is taken  on $X$ while the amount  $(1-\beta)L_t$ is kept in the money market account at  the risk-free rate $r_t$. As a result, the  LETF price satisfies 
\begin{equation*}
\begin{aligned}
\frac{dL_t}{L_t}
&=\beta\left(\frac{dX_t}{X_t}\right)-((\beta-1)r_t)\,dt\\
&=(\beta \mu_t-(\beta-1)r_t)\,dt+\beta\sigma_t\cdot dB_t\;.
\end{aligned}
\end{equation*}
Without loss of generality, we set $L_0=X_0=1.$

The LETF value at time $t$ admits the expression 
 \begin{align}
L_t
&=X_t^\beta e^{\int_0^t(-(\beta-1)r_s -\frac{1}{2}\beta(\beta-1) |\sigma_s|^2)\,ds}\label{lxxt}\\
&=e^{{\int_0^t(\beta\mu_s-(\beta-1)r_s-\frac{1}{2}\beta^2|\sigma_s|^2)\,ds+\beta\int_0^t\sigma_s\cdot\,dB_s}}\;,
\end{align}
  where  $|\,\cdot\,|$ is the usual $d$-dimensional norm. 

 The investor's risk preference is modeled by the power utility function  
\[u(w)={w}^{\alpha}\,, \quad \text{ for } w > 0, \quad \text{ with } \;0<\alpha\leq 1.\]
As such, the  coefficient of relative risk aversion is given by $\varrho:=- w u''(w) / u'(w) = 1-\alpha$, $\forall w>0$. 
The  expected utility from holding the LETF up to time $t$ is given by 
\begin{align}
\mathbb{E}^\mathbb{P}[L_t^\alpha]
&=\mathbb{E}^\mathbb{P}[X_t^{\alpha\beta} e^{\int_0^t(-\alpha(\beta-1)r_s-\frac{1}{2}\alpha\beta(\beta-1)|\sigma_s|^2)\,ds}] \label{EL1}\\
&=\mathbb{E}^\mathbb{P}[e^{\int_0^t(\alpha\beta\mu_s -\alpha(\beta-1) r_s-\frac{1}{2}\alpha\beta^2 |\sigma_s|^2)\,ds+\alpha\beta\int_0^t\sigma_s\cdot\,dB_s}]\label{eqn:expected_utility_of_LETF}\\
&=\mathbb{E}^\mathbb{P}[H_t\,e^{\int_0^t(\alpha\beta\mu_s -\alpha(\beta-1) r_s-\frac{1}{2}\alpha(1-\alpha)\beta^2 |\sigma_s|^2)\,ds}]\;,\label{EL3}
\end{align} 
where we have defined the stochastic exponential
\begin{align}\label{Hproc}
H_t:=e^{\alpha\beta\int_0^t\sigma_s\cdot\,dB_s-\frac{1}{2}\alpha^2\beta^2\int_0^t|\sigma_s|\,ds}\,.\end{align}
In particular, when $\alpha = 1$, the risk aversion $\varrho$ is zero so that the expectation \eqref{EL1} is  the expected return from holding the LETF $L$ over $[0,t]$.

Suppose that a local martingale  $H_t$ in \eqref{Hproc} 
is a martingale. Then, we can define a new measure $\hat{\mathbb{P}}$  via  
\begin{align}
 \frac{d\mathbb{\hat{\mathbb{P}}}}{d\mathbb{P}}\Big|_{\mathcal{F}_t}=H_t\;.\label{dphatdp}
\end{align}
By   Girsanov theorem, the process  $\hat{B}$ defined by
\begin{equation}\label{eqn:BM_under_hat_P}
\hat{B}_t:=-\alpha\beta\int_0^t\sigma_s\,ds+B_t\quad \textnormal {for } \;\;t\ge 0
\end{equation} 
is a standard Brownian motion under $\mathbb{\hat{\mathbb{P}}}.$
Applying \eqref{eqn:BM_under_hat_P} to \eqref{dXt1} and \eqref{EL3}, we get 
$$\frac{dX_t}{X_t}=(\mu_t+\alpha\beta|\sigma_t|^2)\,dt+\sigma_t\cdot d\hat{B}_t$$
and 
\begin{equation}\label{eqn:expected_utility_of_LETF_hat_P}
\begin{aligned}
&\mathbb{E}^\mathbb{P}[L_t^\alpha]
=\mathbb{E}^\mathbb{\hat{\mathbb{P}}}\left[e^{\int_0^t(\alpha\beta\mu_s -\alpha(\beta-1) r_s-\frac{1}{2}\alpha(1-\alpha)\beta^2 |\sigma_s|^2)\,ds}\right]\;.
\end{aligned}
\end{equation}
To analyze the expected utility, we employ the {\em martingale extraction method}, which will be described in Section \ref{sec:martingale_extraction}.
This method allows us to express the expected utility in a form that is more amenable for analysis and computation.

\subsection{Martingale extraction}
\label{sec:martingale_extraction}
We now discuss  the martingale extraction method  with  a generic multi-dimensional time-homogeneous Markov diffusion process   $G_t$ on $(\Omega,\mathcal{F},\mathbb{P})$ with drift $b(G_t)$ and volatility $\sigma(G_t).$ In the SDE form,
we can write by 
$$dG_t=b(G_t)\,dt+v(G_t)\,dB_t\;,$$
where $b$ is $d$-dimensional column vector and $v$ is a $d\times d$ matrix.  The components of $b$ and $\sigma$ are differentiable functions and assume that the SDE has a strong solution.

The $d$-dimensional  process $G_t$ may represent multiple components of the model, such as the reference, stochastic volatility,   stochastic interest rate, or other stochastic factors.   Fix a continuously differentiable multi-variate function $k(\cdot)$.
Denote  by $\mathcal{L}$ the infinitesimal generator  of $G_t$ with killing rate $k$. 
Suppose that 
$(\lambda,\phi)$ is an eigenpair corresponding to 
\begin{align}\label{Lplp}\mathcal{L}\phi=-\lambda\phi\,,
\end{align} where $\lambda\in\mathbb{R}$ and  $\phi$  is a positive continuous twice-differentiable  function. 
It can be shown that  
\begin{align}\label{Mt}M_t:=e^{\lambda t-\int_0^tk(G_s)\,ds}\,\phi(G_t)\,\phi^{-1}(G_0) \end{align}
is a local martingale by checking that the $dt$-term of $dM_t$ is zero. Refer to \cite{hurd2008explicit} for a relevant topic.
\begin{definition} Let $(\lambda,\phi)$ be an eigenpair
	of  $-\mathcal{L}$ satisfying \eqref{Lplp}.
	When the process $M_t$ defined in equation \eqref{Mt}  is a martingale, we say that the pair  	$(\lambda,\phi)$
	admits the martingale extraction 
	of $e^{-\int_0^tk(G_s)\,ds}.$ In this case,   the eigenpair $(\lambda,\phi)$   is called an {\em admissible} eigenpair.
\end{definition}
In this case, we can express equation \eqref{Mt} as
$$e^{-\int_0^tk(G_s)\,ds}=M_t\,e^{-\lambda t}\,\phi^{-1}(G_t)\,\phi(G_0)\,,$$
and interpret it as the martingale $M_t$ being {\em extracted} from $e^{-\int_0^tk(G_s)\,ds}.$
 With each admissible  eigenpair  $(\lambda,\phi)$, one can define a new measure $\mathbb{Q}^\phi$ by
\begin{align}
\mathbb{Q}^\phi(A):=\int_{A}M_{t}\;d\mathbb{P}
=\mathbb{E}^{\mathbb{P}}\left[\mathbb{I}_{A}
M_{t}\right]\quad\text { for }\; A\in \mathcal{F}_{t}\,.
\label{QA}
\end{align}
This measure $\mathbb{Q}^\phi$ is called the {\em   transformed measure} from $\mathbb{P}$
with respect to $(\lambda,\phi)$. 
For convenience, we use notation $\mathbb{Q}$ instead of $\mathbb{Q}^\phi.$
In turn, we apply a change of measure from $\mathbb{P}$ to $\mathbb{Q}$ to express the expectation
\begin{equation} \label{eqn:operator_decomposition}
\begin{aligned}
\mathbb{E}^{\mathbb{P}}
[e^{-\int_{0}^{t} k(G_{s})ds}f(G_t)]
=\mathbb{E}^{\mathbb{Q}}
[(\phi^{-1}f)(G_t)]\cdot e^{-\lambda t}\,\phi(G_0)
\,.
\end{aligned}
\end{equation}
In many cases, the right-hand side is more amenable to computation and analysis.
For instance, the expectation  $\mathbb{E}^{\mathbb{Q}}[(\phi^{-1}f)(G_t)]$ depends on the marginal distribution of   $G_t$ at time $t$,  whereas $\mathbb{E}^{\mathbb{P}}[e^{-\int_{0}^{t} k(G_s)ds}f(G_t)]$ depends on the whole path of $(G_s)_{0\le s\le t}$. This observation is particularly useful for our analysis of   LETFs since they are also path-dependent. 

The dynamic of $G_{t}$ is also altered under the 
transformed measure $\mathbb{Q}$. To see this, we  define  the {\em Girsanov kernel} associated with  $M_t$ by
\begin{align}\varphi:=v^\top\cdot\frac{\nabla \phi}{\phi}\;, \label{varphi}
\end{align}
then  the martingale  $M_t$ satisfies
\begin{align}\frac{dM_{t}}{M_t}=\varphi(G_t)\,dB_t\,.\label{Mtphi}
\end{align}
According to  the Girsanov theorem, the  process  defined by
\begin{align}\label{WB}W_{t}:=B_{t}-\int_{0}^{t}\varphi(G_{s})\,ds, \qquad t\ge 0,
\end{align}
is a standard Brownian motion under $\mathbb{Q}.$ As a result, given  an  admissible eigenpair $(\lambda,\phi)$, 
 the process $G$ evolves under $\mathbb{Q}$  according to  
	\begin{equation*}	 
	dG_{t}=(b+v\varphi)(G_{t})\, dt+v(G_{t})\, dW_{t} .
	\end{equation*}	
As expected, the eigenfunction $\phi$ arises in the drift adjustment of   $G_t$, but does not affect the diffusion term.

Furthermore, if the density function of $G_t$ under $\mathbb{Q}$ is also available in closed form, one can compute and  analyze 
the expectation   $\mathbb{E}^{\mathbb{Q}}[(\phi^{-1}f)(G_t)].$ 
Not all but for many cases, we will choose an admissible eigenpair such that the term $\mathbb{E}^{\mathbb{Q}}[(\phi^{-1}f)(G_t)]$ converges to a non-zero constant. From this we derive  the long-term growth rate of the expected utility of LETFs. 
\begin{proposition}\label{prop1}
	Let  $(\lambda,\phi)$ be an admissible eigenpair of $\mathcal{L}$, and  $\mathbb{Q}$ be the corresponding transformed measure.
	If  $\mathbb{E}^{\mathbb{Q}}[(\phi^{-1}f)(G_t)]$ converges to a nonzero constant as $t \to \infty$, then   the limit
	\begin{align}
	\lim_{t\rightarrow\infty}\frac{1}{t}\log\mathbb{E}^{\mathbb{P}}
	[e^{-\int_{0}^{t} k(G_{s})ds}f(G_t)]=-\lambda 
	\label{limitt1}
	\end{align} holds.
 \end{proposition}

\section{Univariate processes}
\label{sec:Univari}
We now demonstrate how the martingale extraction technique can be applied to analyze the growth rate of  expected utility for  LETFs. In this section, the reference asset  $X_t$ is a one-dimensional Markov diffusion process      that  satisfies 
\begin{align}\label{dXt}\frac{dX_t}{X_t}=\mu(X_t)\,dt+\sigma(X_t)\,dB_t\;,\quad \;X_0=1\,,\end{align}
  where $B$ is a one-dimensional standard Brownian motion under the subjective measure $\mathbb{P}$. The coefficients   $\mu$ and $\sigma$ are continuously differentiable functions such that  SDE \eqref{dXt}   has a strong solution.   Throughout this section, the short interest rate   is a constant $r>0.$

According to   \eqref{EL1},  the expected utility from  holding the LETF is given by
\begin{equation}\label{eqn:LETF_const_short_rate}
	 \mathbb{E}^\mathbb{P}[L_t^\alpha]=\mathbb{E}^\mathbb{P}[X_t^{\alpha\beta}\,e^{-\frac{1}{2}\alpha\beta(\beta-1)\int_0^t\sigma^2(X_u)\,du}]\,e^{r\alpha(1-\beta)t}\;.
\end{equation}
To utilize  the martingale extraction method, we can view  $X_t$  as playing the role of the process $G_t$ in Section \ref{sec:martingale_extraction}.
Define  $\mathcal{L}$ as the infinitesimal generator of $X_t$ with killing rate
$-\frac{1}{2}\alpha\beta(\beta-1)\sigma^2(\cdot)$. As such, we have 
\begin{align}\label{Lx}\mathcal{L}\phi(x)=\frac{1}{2}x^2\sigma^2(x)\phi''(x)+x\mu(x)\phi'(x)-\frac{1}{2}\alpha\beta(\beta-1)\sigma^2(x)\phi(x)\;. \end{align}
A key step in our approach is to  find, as explicitly as possible, an eigenpair $(\lambda,\phi)$ of
$\mathcal{L}\phi=-\lambda\phi$ with positive $\phi.$
It is noteworthy that there always exists such a solution pair as long as  $\beta(\beta-1)\geq 0$ (see  \citep[Theorem 3.3]{Pinsky1995}). This condition is satisfied for all  LETFs since their leverage ratios satisfy    $\beta \notin [0,1]$.

Given that there exists an eigenpair $(\lambda,\phi)$ which admits the martingale extraction of $e^{-\frac{1}{2}\alpha\beta(\beta-1)\int_0^t\sigma^2(X_s)\,ds}$,  the expected utility can be expressed as  
\begin{equation}\label{eqn:after_trans}
\mathbb{E}^\mathbb{P}[L_t^\alpha]
=\mathbb{E}^\mathbb{Q}[X_t^{\alpha\beta}\,\phi^{-1}(X_t)]\,e^{(r\alpha(1-\beta)-\lambda)t}\phi(1)\;,
\end{equation}
where $\mathbb{Q}$ is the corresponding transformed measure.
Since the term $\mathbb{E}^\mathbb{Q}[X_t^{\alpha\beta}\,\phi^{-1}(X_t)]$ depends only on the   value $X_t$ at time $t$, rather than its whole path, this significantly simplifies the analysis of 
$\mathbb{E}^\mathbb{P}[L_t^\alpha]$, as we present in the following models.

 Applying Proposition \ref{prop1}, we obtain the long-term growth rate of expected utility from holding the LETF in this univariate framework. Precisely, we have  
\begin{align} \label{limitaa}	\lim_{t\rightarrow\infty}\frac{1}{t}\log\mathbb{E}^{\mathbb{P}}[L_t^\alpha]= \lim_{t\rightarrow\infty}\frac{1}{t}\log \mathbb{E}^\mathbb{Q}[X_t^{\alpha\beta}\,\phi^{-1}(X_t)]+r\alpha(1-\beta)-\lambda \,,
\end{align} 
and   if $\mathbb{E}^\mathbb{Q}[X_t^{\alpha\beta}\,\phi^{-1}(X_t)]$ converges to a nonzero constant as $t \to \infty,$ then  the limit in \eqref{limitaa} reduces to 
\begin{align} \label{tlt}	\lim_{t\rightarrow\infty}\frac{1}{t}\log\mathbb{E}^{\mathbb{P}}[L_t^\alpha]= r\alpha(1-\beta)-\lambda \,.
\end{align}
In particular, we recover  the  long-term growth rate of expected return for the LETF by setting  $\alpha=1$ corresponding to zero risk aversion. Again, the eigenvalue plays a crucial role in the long-term growth rate, along with the first term that depends explicitly on the interest rate $r$, risk aversion parameter $\alpha$, and the leverage ratio $\beta$. It is important to note that the eigenvalue $\lambda$ also depends on $\alpha$, $\beta$, $\mu(\cdot)$ and $\sigma(\cdot)$, but not $r$.

\subsection{The GBM model}
As a warm-up exercise, we present the  long-term growth rate of expected utility in  the geometric Brownian motion (GBM) model 
$$dX_t=\mu X_t\,dt+\sigma X_t\,dB_t\,,\;t\geq0$$
with $\sigma\neq0.$
The
corresponding generator is
\begin{align}
\mathcal{L}\phi(x)=\frac{1}{2}\sigma^2x^2 \phi''(x)+\mu x \phi'(x)-\frac{1}{2}\alpha\beta(\beta-1)\sigma^2 \phi(x)\;.\end{align}
To apply martingale extraction, we find the corresponding eigenpair
$$(\lambda,\phi(x))=(-\alpha\beta\mu+\frac{1}{2}\alpha(1-\alpha)\beta^2\sigma^2,x^{\alpha\beta})\,.$$
We obtain the expected utility 
$$\mathbb{E}^\mathbb{P}[L_t^\alpha]=
\mathbb{E}^\mathbb{Q}[1]\,e^{(\alpha\beta\mu-\alpha(\beta-1) r-\frac{1}{2}\alpha(1-\alpha)\beta^2\sigma^2)t}
=e^{(\alpha\beta\mu-\alpha(\beta-1) r-\frac{1}{2}\alpha(1-\alpha)\beta^2\sigma^2)t}\;.$$
This implies   the limit
\begin{equation}\label{GBMlambda}
\lim_{t\rightarrow\infty}\frac{1}{t}\log \mathbb{E}^\mathbb{P}[L_t^\alpha]= \alpha(1-\beta) r + \alpha\beta\mu-\alpha(\beta-1) r-\frac{1}{2}\alpha(1-\alpha)\beta^2\sigma^2\;.
\end{equation}
The right-hand side consists of two parts: the factor $\alpha(1-\beta) r$ and the negative eigenvalue $-\lambda$.  Moreover, the long-term growth rate is quadratic in $\beta$. 
Using this result, we maximize the long-term  growth rate in equation \eqref{GBMlambda} over $\beta\in \mathbb{R}$ to  obtain  the optimal leverage ratio 
\begin{align}\beta^*=\frac{\mu-r}{(1-\alpha)\sigma^2}.\end{align}
As we can see, the optimal leverage ratio is wealth independent, proportional to the Sharpe ratio, but inversely proportional to  the  coefficient of relative risk aversion $\varrho = 1-\alpha$. The investor should select a positive (resp. negative) $\beta^*$  if and only if  $\mu > r$ (resp. $\mu < r$). It  resembles the optimal strategy in the classical Merton portfolio optimization problem.

\subsection{The GARCH model}\label{sect-garch}
In  this  section, we consider  a  positive    mean-reverting model for the reference  price process   $X_t$. Specifically, it satisfies  the continuous-time GARCH diffusion model  (see \cite{Lewis2000}):
\begin{align} \label{modelgarch}
dX_t=(\theta-aX_t)\,dt+\sigma X_t\,dB_t\,,
\end{align}
with $a, \theta, \sigma>0.$  The GARCH diffusion model   is sometimes   referred to as the \emph{inhomogeneous geometric Brownian motion} (see e.g. \cite{Zhao2009}).   
The corresponding generator is
\begin{align} \mathcal{L}\phi(x)=\frac{1}{2}\sigma^2x^2\phi''(x)+(\theta-ax)\phi'(x)-\frac{1}{2}\alpha\beta(\beta-1)\sigma^2\phi(x)\;.\end{align}
To apply   martingale extraction,   we solve the eigenpair problem $\mathcal{L} \phi =- \lambda \phi$ to obtain  the eigenpair
$$(\lambda,\phi(x))=\Big(\frac{1}{2}\alpha\beta(\beta-1)\sigma^2,1\Big)\,.$$ Since the eigenfunction $\phi(x)=1$ is just a constant, the   transformed measure $\mathbb{Q}$ coincides with the original measure $\mathbb{P}$ (see \eqref{varphi}-\eqref{WB}). 
Following from \eqref{eqn:after_trans}, the expected utility is
\begin{align}\label{eplt}\mathbb{E}^\mathbb{P}[L_t^\alpha]
=\mathbb{E}^\mathbb{Q}[X_t^{\alpha\beta}]\,e^{(r\alpha(1-\beta)-\frac{1}{2}\alpha\beta(\beta-1)\sigma^2)t}\;.\end{align}
To evaluate \eqref{eplt}, we first deduce that 
\begin{equation}\label{eqn:GARCH}
\left\{\enspace
\begin{aligned}
\lim_{t\rightarrow\infty}&\mathbb{E}^\mathbb{Q}[X_t^{\alpha\beta}]=\textnormal{(positive constant)}\enspace&&\textnormal{ if }\;\,-\alpha\beta+ \frac{2a}{\sigma^2}+1>0\,,\\
&\mathbb{E}^\mathbb{Q}[X_t^{\alpha\beta}]=\infty &&\textnormal{ otherwise}\;.
\end{aligned}\right.
\end{equation}
The proof is as follows. 
The process $Y_t:=\frac{2\theta}{\sigma^2X_t}$ converges to the Gamma
random variable with parameter $\gamma=\frac{2a}{\sigma^2}+1,$ that is,
the density function $p(y;t)$ of $Y_t$ 
converges to $p(y;\infty):=\frac{1}{\Gamma(\gamma)}y^{\gamma-1}e^{-y}$ 
as $t\rightarrow\infty$ (Theorem 2.5 in \cite{Zhao2009}).
We obtain the above result by considering the density function $p(y;t)$ and the limiting density function $p(y;\infty)$ above. 
The asymptotic behaviors of $p(y,t)$   near $y=0$ and $y=\infty$ are as follows. For fixed $t$ and any small $\epsilon>0,$ 
\begin{equation}
\begin{aligned}
&y^{\frac{2a}{\sigma^2}}\lesssim p(y;t)\lesssim  y^{\frac{2a}{\sigma^2}-\epsilon}  &&\textnormal{as }\; y\to 0\\
&p(y;t)\lesssim e^{(-1+\epsilon)y}  &&\textnormal{as }\; y\to \infty  \,.
\end{aligned}
\end{equation}
Here, for two positive functions $p(y)$ and $q(y),$  we denote by
$$p(y)\lesssim q(y)$$ if there exists a positive constant $c$ such that $p(x)\leq c\cdot q(x).$ 
Refer to Section 6.5.4 in \cite{Linetsky2004} for the density funtion $p(y;t).$ If $-\alpha\beta+ \frac{2a}{\sigma^2}+1>0,$ then 
\begin{equation}\label{eqn:GARCH_finite}
\begin{aligned}
\mathbb{E}^\mathbb{Q}[X_t^{\alpha\beta}]
=\Big(\frac{2\theta}{\sigma^2}\Big)^{\alpha\beta}\,\mathbb{E}^\mathbb{Q}[Y_t^{-\alpha\beta}]
&=\Big(\frac{2\theta}{\sigma^2}\Big)^{\alpha\beta}\int_0^\infty y^{-\alpha\beta}p(y;t)\,dy \\
&\rightarrow
\Big(\frac{2\theta}{\sigma^2}\Big)^{\alpha\beta}\frac{1}{\Gamma(\gamma)}\int_0^\infty y^{-\alpha\beta+\frac{2a}{\sigma^2}}e^{-y}\,dy\;, \quad \text{ as } t\to\infty\,,
\end{aligned} 
\end{equation}
which is finite. Otherwise, 
$$\mathbb{E}^\mathbb{Q}[X_t^{\alpha\beta}]
=\Big(\frac{2\theta}{\sigma^2}\Big)^{\alpha\beta}\,\mathbb{E}^\mathbb{Q}[Y_t^{-\alpha\beta}]
=\Big(\frac{2\theta}{\sigma^2}\Big)^{\alpha\beta}\int_0^\infty y^{-\alpha\beta}p(y;t)\,dy=\infty$$ since $y^{\frac{2a}{\sigma^2}}\lesssim p(y;t)$ near $y=0.$
In conclusion, we obtain the following long-term growth  rate.
\begin{proposition} \label{propgarch}
	Let $L_t$ be the LETF whose  reference  price  $X_t$ satisfies  the  GARCH model  \eqref{modelgarch}.	Then, 
	\begin{equation}\label{limitg}
	\lim_{t\rightarrow\infty}\frac{1}{t}\log \mathbb{E}^\mathbb{P}[L_t^\alpha]
	=\left\{\,
	\begin{aligned}
	&r\alpha(1-\beta)-\frac{1}{2}\alpha\beta(\beta-1)\sigma^2&&\textnormal{ if } \,			 
	\frac{2a}{\sigma^2}+1>\alpha\beta\,,\\
	&\quad\quad\quad\infty&&\textnormal{ otherwise . } 		
	\end{aligned} \right.
	\end{equation}
\end{proposition}

 This result implies two distinct scenarios. When  $\frac{2a}{\sigma^2}+1> \alpha\beta$,   there is a finite long-term limit of the growth rate. Interestingly the long-term limit is linear in  $\alpha$ and decreasing in $\sigma^2$, but does not depend on  $\theta$.  When  $\frac{2a}{\sigma^2}+1\le \alpha\beta$, the long-term limit is infinitely large. The limit also applies when $\alpha = 1$, in which case the condition $\beta <\frac{2a}{\sigma^2}+1$ represents an upper bound on the leverage ratio in order to obtain a finite long-term growth rate of return.

 By Proposition \ref{propgarch} and direct calculation, we maximize 
$$\Lambda(\beta):=r\alpha(1-\beta)-\frac{1}{2}\alpha\beta(\beta-1)\sigma^2$$
to obtain the optimal leverage ratio  $\beta^*$ for a long-term investor
\[\beta^*=\frac{\,1\,}{2}-\frac{r}{\sigma^2}.\]
Surprisingly, in  contrast to the GBM model, the optimal leverage ratio under the GARCH model is  independent of   $\alpha$, which means that under this model investors with different risk aversion coefficients, including zero risk aversion,  will have the same optimal leverage ratio $\beta^*$. In fact, $\beta^*$ only depends on the interest rate $r$ and volatility parameter $\sigma$. It is also notable  that the  GARCH model is reduced to the GBM model as $\theta\rightarrow0$; however, not only the optimal growth rate but also the optimal leverage ratio $\beta^*$ do not converge to those of   the geometric Brownian motion as $\theta\rightarrow0.$ It is because the     path behaviors and other qualitative features  of the GARCH model differ significantly from the  GBM model.

\subsection{The inverse GARCH model}
\label{sec:inv_GARCH}
As an alternative to the GARCH model, suppose now    the reference price  $X_t$ follows the inverse GARCH diffusion model, which is also referred to as the Pearl-Verhulst logistic process in \cite{tuckwell1974study}:
\begin{equation}\label{eqn:univariate_inverse_GARCH}
\begin{aligned}
dX_t=(\theta-a X_t)X_t\,dt+\sigma X_t\,dB_t\,,
\end{aligned}
\end{equation}
with $a,\sigma>0$ and $\theta>\sigma^2.$
Both the GARCH and  inverse GARCH models  are  positive and mean-reverting. The process $X_t$ is called the inverse GARCH model because its inverse process $Y_t:=1/X_t$ follows  the GARCH model:
$$dY_t=(a-(\theta-\sigma^2)Y_t)\,dt-\sigma Y_t\,dB_t\;.$$

The infinitesimal  generator of $X_t$ is 
\begin{align} \mathcal{L}\phi(x)=\frac{1}{2}\sigma^2x^2\phi''(x)+(\theta-ax)x\phi'(x)-\frac{1}{2}\alpha\beta(\beta-1)\sigma^2\phi(x)\;.\end{align}
By direct substitution, we verify that 
$$\left(\lambda,\phi(x)\right)=\Big(\frac{1}{2}\alpha\beta(\beta-1)\sigma^2,1\Big) $$ 
is an admissible eigenpair to $\mathcal{L} \phi  = -\lambda \phi$. Since the eigenfunction $\phi(x)=1$ is a constant, the corresponding transformed measure $\mathbb{Q}$ is identical to the original measure $\mathbb{P}$  (see \eqref{varphi}-\eqref{WB}). 
Following from \eqref{eqn:after_trans}, 
the expected utility is
\begin{equation}\label{eqn:Inv_GARCH}
\mathbb{E}^\mathbb{P}[L_t^\alpha]
=\mathbb{E}^\mathbb{Q}[X_t^{\alpha\beta}]\,e^{(r\alpha(1-\beta)-\frac{1}{2}\alpha\beta(\beta-1)\sigma^2)t}\;. 
\end{equation}
Since $Y_t=1/X_t$ is the GARCH model, we observe from \eqref{eqn:GARCH} that
\begin{equation*} 
\left\{\enspace
\begin{aligned}
\lim_{t\rightarrow\infty}&\mathbb{E}^\mathbb{Q}[X_t^{\alpha\beta}]
=\lim_{t\rightarrow\infty}\mathbb{E}^\mathbb{Q}[Y_t^{-\alpha\beta}]=\textnormal{(positive constant)}\enspace&&\textnormal{ if }\;\,\alpha\beta+ \frac{2\theta}{\sigma^2}>1\,,\\
&\mathbb{E}^\mathbb{Q}[X_t^{\alpha\beta}]=\infty &&\textnormal{ otherwise}\;.
\end{aligned}\right.
\end{equation*}
This leads to the long-term limit summarized as follows.
\begin{proposition}\label{prop:Inv_GAR}
	Let $L_t$ be the LETF whose  reference price   $X_t$  follows  the  inverse GARCH model \eqref{eqn:univariate_inverse_GARCH}. 
	Then,  
	\begin{equation}\label{limitinvg}
	\lim_{t\rightarrow\infty}\frac{1}{t}\log \mathbb{E}^\mathbb{P}[L_t^\alpha]
	=\left\{\,
	\begin{aligned}
	&r\alpha(1-\beta)-\frac{1}{2}\alpha\beta(\beta-1)\sigma^2&&\textnormal{ if } 			 
	\alpha\beta+\frac{2\theta}{\sigma^2}>1\,,\\
	&\quad\quad\quad\infty&&\textnormal{ otherwise . } 		
	\end{aligned} \right.
	\end{equation}
\end{proposition}
Therefore, the long-term growth rate of expected utility can be finite or infinite, depending on the leverage ratio $\beta$, risk aversion parameter $\alpha$, and model parameters  $(\theta,\sigma)$ but \textit{not} $a$.   Interestingly, the limits in \eqref{limitg} and \eqref{limitinvg}, respectively,  for the GARCH and inverse GARCH models are the same, except for the conditions for  the finiteness of the limits.  

 By direct calculation, the  optimal leverage ratio   for the long-term investor is   $\beta^*=\frac{\,1\,}{2}-\frac{r}{\sigma^2}$ when the long-term growth rate is finite. While $\beta^*$ does not depend explicitly on    $\alpha$, but $\alpha$ plays a role in determining the finite/infinite growth rate scenario.    As  $a\rightarrow 0$, the inverse  GARCH model  reduces to  the GBM model. Nevertheless,   the optimal growth rate  and optimal leverage ratio $\beta^*$, being independent of $a$, do not  converge to those in the GBM model as $a\rightarrow 0$. The same  phenomenon  was observed in the  GARCH model case in Section \ref{sect-garch}.

\subsection{The extended CIR model}
\label{sec:CIR}
We now turn to the extended Cox-Ingersoll-Ross (CIR) model  proposed by \cite{cox1985theory}:
\begin{equation}\label{eqn:extended_CIR}
 dX_t=(\theta+\mu X_t)\,dt+\sigma\sqrt{X_t}\,dB_t  \,,
\end{equation}
with parameters $\mu,\sigma>0$ and $\theta\geq\sigma^2.$ This process is a transient process diverging to infinity given  $\mu>0$. The corresponding infinitesimal generator is given by
$$\mathcal{L}\phi(x)=\frac{1}{2}\sigma^2x\,\phi''(x)+(\theta+\mu x)\,\phi'(x)
-\frac{1}{2}\alpha\beta(\beta-1)\sigma^2\frac{1}{x}\,\phi(x)\;.$$
Set 
$$\kappa:=\sqrt{\left(\frac{1}{2}-\frac{\theta}{\sigma^2}\right)^2+\alpha\beta(\beta-1)}+\frac{1}{2}-\frac{\theta}{\sigma^2}
\;.$$
The first square-root term is real  provided  that   $\beta \notin [0,1]$, which holds true for all LETFs. It can be verified by direct substitution  that
$$(\lambda,\phi(x)):=\left(\mu \kappa+\frac{2\theta\mu}{\sigma^2}\,,\,e^{-\frac{2\mu x}{\sigma^2}}x^\kappa\right)$$
is an admissible eigenpair of $\mathcal{L}$ according to equation \eqref{Lplp}. Under the   transformed measure $\mathbb{Q}$ with respect to this eigenpair, the  process $X_t$ follows
\begin{equation}\label{eqn:ext_CIR_Q}
dX_t=(\theta+\kappa\sigma^2-\mu X_t)\,dt+\sigma\sqrt{X_t}\,dW_t\;, 
\end{equation}
where $W_t$ is a $\mathbb{Q}$-Brownian motion.
We note that this is a standard mean-reverting CIR process and the Feller condition is satisfied, thus $0$ is an unattainable boundary.

The expected utility  is given by
\begin{align}\label{expectucir}\mathbb{E}^\mathbb{P}[L_t^\alpha]
=\mathbb{E}^\mathbb{Q}[X_t^{\alpha\beta-\kappa}\,e^{\frac{2\mu}{\sigma^2}X_t}]\,e^{(r\alpha(1-\beta)-\mu \kappa-\frac{2\theta\mu}{\sigma^2})t-\frac{2\mu}{\sigma^2}}\;.\end{align}
For the RHS of \eqref{expectucir}, we obtain the long-term   limit  (see Appendix \ref{app:the_mart_extrac_CIR}): 
\begingroup\makeatletter\def\f@size{12}\check@mathfonts
\begin{equation}\label{eqn:ext_CIR_lim}
\lim_{t\rightarrow\infty}\frac{1}{t}\log \mathbb{E}^\mathbb{Q}[X_t^{\alpha\beta-\kappa}e^{\frac{2\mu}{\sigma^2}X_t}]
=\left\{
\begin{aligned}
&\Big(\alpha\beta+\frac{2\theta}{\sigma^2}+\kappa\Big)\mu\quad&&\textnormal{ if }\;\alpha\beta+\frac{2\theta}{\sigma^2}+\kappa>0\;,\\
&\quad\quad\infty&&\textnormal{ if }\; \alpha\beta+\frac{2\theta}{\sigma^2}+\kappa\leq0\;.\\
\end{aligned}\right.
\end{equation} 
\endgroup
In turn, we obtain the long-term growth rate of expected utility.
\begin{proposition}
Suppose that the reference price process 	$X_t$ satisfies  the extended CIR model \eqref{eqn:extended_CIR}.  Then, we have 
\begin{equation*}
\begin{aligned}
\lim_{t\rightarrow\infty}\frac{1}{t}\log \mathbb{E}^\mathbb{P}\left[L_t^\alpha\right] 
=\;\left\{
\begin{aligned}
\alpha r+&\alpha\beta(\mu-r)&&\textnormal{ if }\;\alpha\beta+\frac{2\theta}{\sigma^2}+\kappa>0\;,\\
&\infty &&\textnormal{ if }\;\alpha\beta+\frac{2\theta}{\sigma^2}+\kappa\leq0\;.\\
\end{aligned}\right.
\end{aligned} 
\end{equation*} 
\end{proposition}

This result has a number of implications.  First, the long-term growth rate is affine in the leverage ratio $\beta$ and excess return $(\mu-r)$, and linear in $\alpha$, but it does not depend on the model parameters $\theta$ and $\sigma$ explicitly other than in the   condition separating the two scenarios. In the scenario with 
$\alpha\beta+\frac{2\theta}{\sigma^2}+\kappa>0$,  
denote the limit as a function of $\beta$: $\Lambda(\beta)
:=\alpha r+\alpha\beta(\mu-r)$.  When $\beta=0$, it follows that the  long-term growth rate $\Lambda (0) = \alpha r$. This is because the resulting ``leveraged" ETF portfolio is simply growing deterministically at rate $r$, and the utility is $e^{\alpha rt}$ at time $t\ge 0$. 
Second, the function $\Lambda(\beta)$ reveals the optimal choice $\beta^*$ for a static investor.
In a bullish market with $\mu>r$, a  higher  leverage ratio  is   preferred, though  in practice  the available  leverage ratios are capped at $+3$. 
In contrast, in a bearish market with $\mu<r,$ then  a  more negative  leverage ratio   is better, and in practice the  most negative leverage ratio  available among LETFs is  $-3$.

\subsection{The 3/2 model}
\label{sec:3/2_model}
We now consider the 3/2 model     for the reference price $X_t$   of the form:
\begin{equation}\label{eqn:32}
dX_t=(\theta-aX_t)X_t\,dt+\sigma X_t^{3/2}\,dB_t\,,
\end{equation}
with $a, \theta, \sigma>0.$ This is a  positive mean-reverting model that   has been used to model  interest rates and volatility (see \cite{ahn1999parametric}, \cite{carr2007new}), so this model would be appropriate for fixed-income and volatility LETFs with a mean-reverting reference price.

The   infinitesimal generator corresponding to \eqref{eqn:32} is 
$$\mathcal{L}\phi(x)=\frac{1}{2}\sigma^2x^3\,\phi''(x)+(\theta-ax)x\,\phi'(x)-\frac{1}{2}\alpha\beta(\beta-1)\sigma^2x\,\phi(x)\;.$$
Denoting  $$\kappa:=\sqrt{\left(\frac{1}{2}+\frac{a}{\sigma^2}\right)^2+\alpha\beta(\beta-1)}-\left(\frac{1}{2}+\frac{a}{\sigma^2}\right)\;,$$
we find that 
$$(\lambda,\phi(x)):=\left(\theta\kappa\,,\,x^{-\kappa}\right)$$
is an admissible eigenpair of $\mathcal{L}.$ Under the transformed measure $\mathbb{Q}$, the reference price $X_t$ follows
$$dX_t=(\theta -(a+\sigma^2\kappa)X_t )X_t\,dt+\sigma X_t^{3/2}\,dW_t\;,$$
 where $dW_t =dB_t+\sigma\kappa X_t^{1/2} dt$ is a Brownian motion under $\mathbb{Q}.$ Notice that $X_t$ satisfies a  re-parametrized $3/2$ model under $\mathbb{Q}$. 

The expected utility from  holding an LETF can be expressed under the transformed measure $\mathbb{Q}$ by
$$\mathbb{E}^\mathbb{P}[L_t^\alpha]
=\mathbb{E}^\mathbb{Q}[X_t^{\alpha\beta+\kappa}]\,e^{(r\alpha(1-\beta)-\theta\kappa)t}\;.$$
We show  that  
\begin{equation*}\left\{\enspace
\begin{aligned}
\lim_{t\rightarrow\infty}&\mathbb{E}^\mathbb{Q}[X_t^{\alpha\beta+\kappa}]=\textnormal{(positive constant)}\enspace&&\textnormal{ if }\;\,  \frac{2a}{\sigma^2}+\kappa-\alpha\beta+2>0\,,\\
&\mathbb{E}^\mathbb{Q}[X_t^{\alpha\beta+\kappa}]=\infty &&\textnormal{ otherwise}\;.
\end{aligned}\right.
\end{equation*}
The proof is as follows.
Define $Y_t:=1/X_t.$ Then $Y_t$ is a CIR process with
$$dY_t=(a+\sigma^2(\kappa+1)-\theta Y_t)\,dt-\sigma\sqrt{Y_t}\,dW_t\;.$$
By considering the density function of the CIR process, which is given in equation \eqref{eqn:CIR_density},
we obtained the desired result. 
	In conclusion, we obtain the following long-term growth  rate.
\begin{proposition}
		Let $L_t$ be a $\beta$-LETF whose   reference price    $X_t$ satisfies  the   $3/2$ model \eqref{eqn:32}. Then, we have 
		\begin{equation*}
		\lim_{t\rightarrow\infty}\frac{1}{t}\log \mathbb{E}^\mathbb{P}\left[L_t^\alpha\right]
		=\left\{\,
		\begin{aligned}
		&r\alpha(1-\beta)-\theta\kappa&&\textnormal{ if  } 			 
		\frac{2a}{\sigma^2}+\kappa-\alpha\beta+2>0\,,\\
		&\quad\quad\quad\infty&&\textnormal{ otherwise\,.} 		
		\end{aligned} \right.
		\end{equation*}	 

\end{proposition}
\noindent In general, the sign of $\frac{2a}{\sigma^2}+\kappa-\alpha\beta+2$ depends on the model parameters $(\theta, a, \sigma)$, risk  aversion coefficient $\alpha$,  and leverage ratio $\beta$. Nevertheless,  we find that for $|\beta|\leq 3$, which holds for market-traded LETFs, the condition $\frac{2a}{\sigma^2}+\kappa-\alpha\beta+2>0$ is satisfied.

Next, we investigate the optimal leverage ratio  $\beta^*$ for a static investor (see \eqref{maxbeta}). In the scenarios with  $\frac{2a}{\sigma^2}+\kappa-\alpha\beta+2>0$, we define
$$\Lambda(\beta):=r\alpha(1-\beta)-\theta\kappa\;.$$
Next, we determine the critical points of $\Lambda.$
Differentiation yields    that 
$$\Lambda'(\beta)=-r\alpha-\frac{\theta\alpha(2\beta-1)}{2\sqrt{\left(\frac{1}{2}+\frac{a}{\sigma^2}\right)^2+\alpha\beta(\beta-1)}}\;.$$
When $\alpha\geq\frac{\theta^2}{r^2},$ the equation $\Lambda'(\beta)=0$ has no solutions and $\Lambda'(\beta)<0$ for all $\beta$. Therefore,   $\Lambda(\beta)$ is a decreasing function of $\beta.$ In practice, $\beta^*=-3$ is the optimal strategy. 
On the other hand, when  $\alpha<\frac{\theta^2}{r^2},$
by considering the equation
$\Lambda'(\beta)=0,$
we conclude that the maximum of $\Lambda(\beta)$ is attained  at
$$\beta^*=\frac{1}{2}-\frac{1}{2}\sqrt{\frac{\left(1+\frac{2a}{\sigma^2}\right)^2 -\alpha}{\frac{\theta^2}{r^2}-\alpha}}\;.$$
Note that the number inside the square root is positive because $\alpha\leq1<(1+ {2a}{\sigma^{-2}})^2.$
Moreover, the  optimal value satisfies
 \[\beta^*=0 \qquad \text{ if and only if } \qquad 1+\frac{2a}{\sigma^2}=\frac{\theta}{r}.\]

\section{Stochastic volatility models}
\label{sec:SV}
In this section, we analzye  the martingale extraction method for LETFs under stochastic volatility models.
Let $B_t$ be
a standard  Brownian motion under $\mathbb{P}.$
The reference price $X_t$ satisfies the SDE 
$$\frac{dX_t}{X_t}=\mu\,dt+\sigma(Y_t)\cdot dB_t$$
with a constant $\mu,$ a $d$-dimensional column vector  $\sigma(\cdot)$ 
and a Markov diffusion process $Y_t$ as the driver of the stochastic volatility. 
Throughout this section, the   interest rate  is a constant $r>0$.

As discussed in Section \ref{sec:LETF_dynamics},
define a new measure  $\hat{\mathbb{P}}$  by $$ \frac{d\mathbb{\hat{\mathbb{P}}}}{d\mathbb{P}}\Big|_{\mathcal{F}_t}=e^{\alpha\beta\int_0^t\sigma(Y_s) dB_s-\frac{1}{2}\alpha^2\beta^2\int_0^t|\sigma|^2(Y_s)\,ds}\;.$$
Then, the process defined   by
\begin{equation*} 
\hat{B}_t:=-\alpha\beta\int_0^t\sigma(Y_s)\,ds+B_t
\end{equation*} 
is a standard  Brownian motion under $\mathbb{\hat{\mathbb{P}}}.$
From equation  \eqref{eqn:expected_utility_of_LETF_hat_P}, it follows that  
\begin{equation*}
\begin{aligned}
&\mathbb{E}^\mathbb{P}[L_t^\alpha]=\mathbb{E}^\mathbb{\hat{\mathbb{P}}}[e^{-\frac{1}{2}\alpha(1-\alpha)\beta^2\int_0^t|\sigma|^2(Y_s)\,ds}]\,e^{\alpha(r+\beta(\mu-r))t}\;.
\end{aligned}
\end{equation*}
Following \eqref{eqn:operator_decomposition} with the stochastic volatility driver  $Y_t$ here playing the role of $G_t$ in Section \ref{sec:martingale_extraction}, the  main idea is to apply the martingale extraction of
$e^{-\frac{1}{2}\alpha(1-\alpha)\beta^2\int_0^t|\sigma|^2(Y_s)\,ds}$  and compute the expected utility explicitly.

\subsection{The Heston model}
\label{sec:Heston}
We now present an  example in which the reference price follows  the Heston model (see \cite{heston1993closed})  
\begin{equation}\label{eqn:Heston}
\begin{aligned}
&dX_t=\mu X_t\, dt+ \sqrt{v_t}X_t\,dB_t\;,\\
&dv_t=(\theta-a v_t)\,dt+\delta\sqrt{v_t}\,dZ_t\;,
\end{aligned}
\end{equation}
where $B_t$ and $Z_t$ are two correlated  Brownian motions with  $\langle Z,W\rangle_t=\rho t$  and correlation parameter  $\rho \in [-1,1]$.  This model assumes  that $\mu,\theta,a,\delta>0$ and $2\theta>\delta^2.$

Define the measure $\hat{\mathbb{P}}$ 
via \eqref{dphatdp}-\eqref{eqn:BM_under_hat_P} in Section \ref{sec:LETF_dynamics} so that
the process defined by 
$$\hat{B}_t=-\alpha\beta\int_0^t\sqrt{v_s}\,ds+dB_t$$
is a $\hat{\mathbb{P}}$-Brownian motion. By  \eqref{eqn:expected_utility_of_LETF_hat_P}, we express the expected utility under  the   measure $\hat{\mathbb{P}}$ as
\begin{equation*}
\begin{aligned}
\mathbb{E}^\mathbb{P}[L_t^\alpha]=\mathbb{E}^\mathbb{\hat{\mathbb{P}}}[e^{-\frac{1}{2}\alpha(1-\alpha)\beta^2\int_0^tv_s\,ds}]\,e^{\alpha(r+\beta(\mu-r))t}\;.
\end{aligned}
\end{equation*}
The stochastic volatility process  $v_t$ is  a re-parametrized CIR process 
\begin{equation*}
\begin{aligned}
dv_t
&=(\theta-(a-\alpha\beta\delta\rho) v_t)\,dt+\delta\sqrt{v_t}\,d\hat{Z}_t\,,
\end{aligned}
\end{equation*}
where  $\hat{Z}_t$ is another Brownian motion under  $\hat{\mathbb{P}}$.

We now explore the martingale extraction of
$e^{-\frac{1}{2}\alpha(1-\alpha)\beta^2\int_0^tv_s\,ds}$. To this end, we consider stochastic volatility process  $v_t$  as playing  the role of the process $G_t$ discussed in Section \ref{sec:martingale_extraction}.
 The infinitesimal generator $\mathcal{L}$ of     $v_t$ with killing rate
$\frac{1}{2}\alpha(1-\alpha)\beta^2v_t$
is
$$\mathcal{L}\phi(v)=\frac{1}{2}\delta^2v\,\phi''(v)+(\theta-(a-\alpha\beta\delta\rho) v)\,\phi'(v)-\frac{1}{2}\alpha(1-\alpha)\beta^2v\,\phi(v)\;.$$
By direct calculation, we obtain  an admissible eigenpair of $\mathcal{L}$, given by 
\begin{equation*}
(\lambda,\phi(v))=\left(\theta\kappa,e^{-\kappa v}\right) \,,
\end{equation*}
where
\[\kappa:=\frac{1}{\delta^2}(\sqrt{(a-\alpha\beta\delta\rho)^2+\alpha(1-\alpha)\beta^2\delta^2}-a+\alpha\beta\delta\rho)\;.\]
Let $\mathbb{Q}$ be the transformed measure with respect to this pair $(\lambda,\phi(v)).$ Then, the process $v_t$ satisfies another re-parametrized CIR model 
$$dv_t=(\theta-\sqrt{(a-\alpha\beta\delta\rho)^2+\alpha(1-\alpha)\beta^2\delta^2}\,v_t)\,dt+\delta\sqrt{v_t}\,dW_t\;,$$
where $W_t$ is a $\mathbb{Q}$-Brownian motion. 
The expected utility can be written as  
\[\mathbb{E}^\mathbb{P}[L_t^\alpha]=\mathbb{E}^\mathbb{\hat{\mathbb{P}}}[e^{-\frac{1}{2}\alpha(1-\alpha)\beta^2\int_0^tv_s\,ds}]\,e^{\alpha(r+\beta(\mu-r))t}
=\mathbb{E}^\mathbb{Q}[e^{\kappa v_t}]\,e^{(\alpha r+\alpha\beta(\mu-r)-\theta\kappa) t-\kappa v_0}.\]
Note that 
$\mathbb{E}^\mathbb{Q}[e^{\kappa r_t}]$ converges to a nonzero constant as $t\rightarrow\infty.$
We refer  to Appendix \ref{app:the_mart_extrac_CIR} for more details about this martingale extraction. 
\begin{proposition}\label{propheston}
		Let $L_t$ be the LETF whose   reference price     $X_t$    satisfies  the Heston model \eqref{eqn:Heston}.
		Then, we have 
\begin{align}
\lim_{t\rightarrow\infty}\frac{1}{t}\log \mathbb{E}^\mathbb{P}&[L_t^\alpha]=\alpha r+\alpha\beta(\mu-r)-\theta\kappa\notag\\
&=\alpha r+\alpha\beta(\mu-r)-\frac{\theta}{\delta^2}(\sqrt{(a-\alpha\beta\delta\rho)^2+\alpha(1-\alpha)\beta^2\delta^2}-a+\alpha\beta\delta\rho).\label{limitheston}
\end{align}	
\end{proposition}

\begin{figure}\begin{center} \includegraphics[trim={0cm 0cm 0cm 0cm},clip,scale = 0.6]{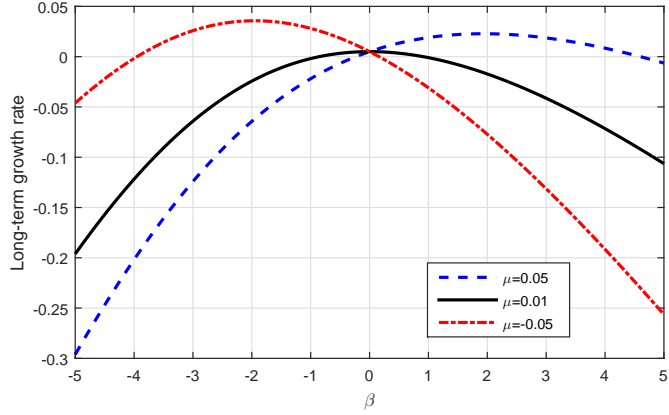} \caption{Long-term growth rate of expected utility as a function of the leverage ratio $\beta$ under the  Heston model. Under three different   values  of the drift  $\mu\in\{0.05, 0.01, -0.05\}$, the  optimal $\beta^*$  maximizing   the  growth rates  are $\{1.93,  0, -1.95\},$  respectively. } \label{fig_hest}\end{center} \end{figure}

We now determine    the optimal leverage ratio $\beta^*$ for the risk-averse  static investor.
To understand the dependence of the long-term limit in \eqref{limitheston} on $\beta$, we define the function 
$$\Lambda(\beta):=
\left(\frac{\alpha\delta^2(\mu-r)}{\theta}-\alpha\delta\rho \right)\beta-\sqrt{(a-\alpha\beta\delta\rho)^2+\alpha(1-\alpha)\beta^2\delta^2}\,.$$
Let 
\begin{align}
C_1=\alpha(1-\alpha)\delta^2+\alpha^2\delta^2\rho^2\,,\,C_2=-a\alpha\delta\rho\,,\,C_3=a^2\,,\,D=\frac{\alpha\delta^2(\mu-r)}{\theta}-\alpha\delta\rho\;.\label{ccd}\end{align}
Then, we rewrite $\Lambda(\beta)$  to highlight the dependence on $\beta$ as
$$\Lambda(\beta)=D\beta-\sqrt{C_1\beta^2+2C_2\beta+C_3}\,.$$
In turn, we obtain the derivatives:
$$\Lambda'(\beta)=D-\frac{C_1\beta+C_2}{\sqrt{C_1\beta^2+2C_2\beta+C_3}}\;,\;\Lambda''(\beta)=\frac{C_2^2-C_1C_3}{(C_1\beta^2+2C_2\beta+C_3)^{3/2}}\;.$$
Since $C_2^2-C_1C_3<0,$ we know that $\Lambda(\beta)$ is a strictly concave function of $\beta.$ 
\begin{itemize}[noitemsep]
	\item[(i)] If $C_1>D^2,$ then $\Lambda'(\beta)=0$ has a unique  solution, which gives the optimal  leverage ratio
	\begin{align*}
	\beta^*=-\frac{C_2}{C_1}+\frac{|D|}{C_1}\sqrt{\frac{C_1C_3-C_2^2}{C_1-D^2}}\;.
	\end{align*}
	\item[(ii)] If $C_1\leq D^2,$ then $\Lambda'(\beta)=0$ has no solutions. 	Furthermore, if  $D>0,$  $\Lambda'(\beta)$ is positive for all $\beta,$ thus $\Lambda(\beta)$ is an increasing function. The optimal $\beta^*=3$ (or the maximum available leverage ratio) in practice. If  $D<0,$ $\Lambda'(\beta)$ is negative for all $\beta,$ thus $\Lambda(\beta)$ is a decreasing function and the most negative leverage ratio is preferred. In practice, the investor would select    $\beta^*=-3$. This is intuitive since $D<0$ means that $\mu <r$ (see \eqref{ccd}). 
\end{itemize}
Figure \ref{fig_hest} depicts  the long-term growth rate  in Proposition \ref{propheston} as a function of $\beta$. The parameters are: $\alpha=0.5,$   $r=0.01,$ $\theta=0.16,$ $\delta=0.89,$ $
a=3.1,$ $\rho=-0.5$, along with    $\mu\in\{0.05, 0.01, -0.05\}$. As we can see, when the excess return $(\mu-r)$ is positive, then the optimal leverage ratio is positive ($\beta^* = 1.93$ when $\mu -r = 0.04$). In contrast, when the reference price $X_t$ is trending downward ($\mu =-0.05$), then it is optimal for the investor to select a short LETF (i.e. $\beta =-1.95$).

\subsection{The 3/2 volatility model}
Under the  3/2 volatility model proposed by  \cite{carr2007new}, the  reference price $X_t$ follows
\begin{equation}\label{eqn:3/2_vol}
\begin{aligned}
&dX_t=\mu X_t\, dt+ \sqrt{v_t}X_t\,dB_t\;,\\
&dv_t=(\theta-a v_t)v_t\,dt+\delta v_t^{3/2}\,dZ_t\;,
\end{aligned}
\end{equation}
where $B_t$ and $Z_t$ are two  standard  Brownian motions with instantaneous correlation   $\rho \in[-1,1]$. 
 
As discussed in Section \ref{sec:LETF_dynamics}, we define the  measure $\hat{\mathbb{P}}$  so that the process  
$$\hat{B}_t=-\alpha\beta\int_0^t\sqrt{v_s}\,ds+dB_t$$
is a  standard Brownian motion under $\hat{\mathbb{P}}$.
As a result of \eqref{eqn:expected_utility_of_LETF_hat_P}, the  expected utility admits the expression   
\begin{equation*}
	\begin{aligned}
		\mathbb{E}^\mathbb{P}[L_t^\alpha]=\mathbb{E}^\mathbb{\hat{\mathbb{P}}}[e^{-\frac{1}{2}\alpha(1-\alpha)\beta^2\int_0^tv_s\,ds}]\,e^{\alpha(r+\beta(\mu-r))t}\;.
	\end{aligned}
\end{equation*}
The stochastic volatility process    $v_t$  follows a re-parametrized 3/2 model 
\begin{equation*}
\begin{aligned}
dv_t
=(\theta-(a-\alpha\beta\delta\rho) v_t)v_t\,dt+\delta v_t^{3/2}\,d\hat{Z}_t
\end{aligned}
\end{equation*}
where  $\hat{Z}_t$ is a     $\hat{\mathbb{P}}$-Brownian motion.

We apply the   martingale extraction method by viewing the stochastic volatility process  $v_t$ as the process $G_t$ in Section \ref{sec:martingale_extraction}. The infinitesimal generator $\mathcal{L}$ of the diffusion $v_t$ with killing rate $\frac{1}{2}\alpha(1-\alpha)\beta^2v_t$ is
$$\mathcal{L}\phi(v)=\frac{1}{2}\delta^2v^3\,\phi''(v)
+(\theta-(a-\alpha\beta\delta\rho)v)v\,\phi'(v)-\frac{1}{2}\alpha(1-\alpha)\beta^2v\,\phi(v)\;.$$
It can be shown that 
\begin{equation*}
(\lambda,\phi(v)):=\left(\theta\kappa\,,\,v^{-\kappa}\right)
\end{equation*}
is an admissible eigenpair of $\mathcal{L},$
where
$$\kappa:=\frac{1}{\delta^2}(\sqrt{(a-\alpha\beta\delta\rho+\delta^2/2)^2+\alpha(1-\alpha)\beta^2\delta^2}-(a-\alpha\beta\delta\rho+\delta^2/2))\;.$$
Let $\mathbb{Q}$ be the corresponding transformed measure.
The process $v_t$ satisfies
$$dv_t=(\theta-(\sqrt{(a-\alpha\beta\delta\rho+\delta^2/2)^2+\alpha(1-\alpha)\beta^2\delta^2}-\delta^2/2) v_t)v_t\,dt+\delta v_t^{3/2}\,dW_t\;,$$
where $W_t$ is a $\mathbb{Q}$-Brownian motion.  Consequently, we express the expected utility as
$$\mathbb{E}^\mathbb{P}[L_t^\alpha]=\mathbb{E}^\mathbb{\hat{\mathbb{P}}}[e^{-\frac{1}{2}\alpha(1-\alpha)\beta^2\int_0^tv_s\,ds}]\,e^{\alpha(r+\beta(\mu-r))t}
=\mathbb{E}^\mathbb{Q}[v_t^\kappa]\,e^{(\alpha r+\alpha\beta(\mu-r)-\theta\kappa) t}v_0^{-\kappa} \;.$$ 
We show  that
$\mathbb{E}^\mathbb{Q}[v_t^\kappa]$
converges to a positive constant by considering the density of the CIR process $1/v_t$ given that  
$$\frac{1}{\delta^2}(\sqrt{(a-\alpha\beta\delta\rho+\delta^2/2)^2+\alpha(1-\alpha)\beta^2\delta^2}+(a-\alpha\beta\delta\rho+\delta^2/2))+1>0\;.$$
Otherwise, we have $\mathbb{E}^\mathbb{Q}[v_t^\kappa]=\infty.$ In conclusion, we have the following proposition.
\begin{proposition} 
	Let $L_t$ be the LETF with reference process $X_t$ satisfying the  3/2 model  \eqref{eqn:3/2_vol}. 
	Then, we have 
	\begin{align}
	\lim_{t\rightarrow\infty}\frac{1}{t}\log \mathbb{E}^\mathbb{P}[L_t^\alpha]
	=\;\alpha r+\alpha\beta(\mu-r)-\theta\kappa\,,\label{limit32}
	\end{align}	 
	if $$\frac{1}{\delta^2}(\sqrt{(a-\alpha\beta\delta\rho+\delta^2/2)^2+\alpha(1-\alpha)\beta^2\delta^2}+(a-\alpha\beta\delta\rho+\delta^2/2))+1>0\;.$$ 
	Otherwise, we have $$\lim_{t\rightarrow\infty}\frac{1}{t}\log \mathbb{E}^\mathbb{P}[L_t^\alpha]=\infty\;.$$ 
\end{proposition}

The explicit long-term limit allows us to determine conveniently the  optimal leverage ratio $\beta^*$ for the risk-averse  static investor.
To this end, we   define the following function out of \eqref{limit32}
$$\Lambda(\beta):=
((\mu-r)\delta/\theta-\rho)\alpha\delta\beta-\sqrt{(a-\alpha\beta\delta\rho+\delta^2/2)^2+\alpha(1-\alpha)\beta^2\delta^2}\;.$$
We define  the constants \begin{equation*}
\begin{aligned}
&C_1=\alpha(1-\alpha)\delta^2+\alpha^2\delta^2\rho^2\;,\\
&C_2=-\alpha\delta\rho(a+\delta^2/{2})\;,\\
&C_3=(a+\delta^2/{2})^2\;,\\
&D=((\mu-r)\delta/\theta-\rho)\alpha\delta\;, 
\end{aligned}
\end{equation*}
and to highlight its dependence on $\beta$  we   rewrite $\Lambda(\beta)$ as 
$$\Lambda(\beta)=D\beta-\sqrt{C_1\beta^2+2C_2\beta+C_3}\,.$$
Given the same structure of $\Lambda$, the optimal $\beta^*$ can be derived by the exactly same way as  in Section \ref{sec:Heston}.
In summary, if  $C_1>D^2,$ then the optimal $\beta^*$ is
	\begin{align*}
	\beta^*=-\frac{C_2}{C_1}+\frac{|D|}{C_1}\sqrt{\frac{C_1C_3-C_2^2}{C_1-D^2}}\;.
	\end{align*}
If  $C_1\leq D^2$, then it is  optimal to pick the most positive (resp. most negative) $\beta$ possible  if $D>0$ (resp.  $D<0$).

\section{LETF with stochastic reference and interest rate}
\label{sec:interest_rate}
In this section, we analyze  the long-term growth rate of the expected utility from holding an LETF when both the reference price $X_t$  and        short interest rate $r_t$ are stochastic.

\subsection{Vasicek interest rate}
\label{sec:Vasicek}
We first consider  the Vasicek interest rate model introduced by  \cite{vasicek1977equilibrium}.
The reference price process $X_t$ and the short interest rate $r_t$ satisfy the SDEs
\begin{align}
&dX_t=\mu X_t\,dt+\sigma X_t\,dB_t\;,\label{eqn:Vasicek}\\
&dr_t=(\theta-a r_t)\,dt+\delta \,dZ_t\;,\label{eqn:Vasicek2}
\end{align}
for $\mu,\sigma,\theta,a,\delta>0,$
where $B_t$ and $Z_t$ are two   Brownian motions such that  $\langle Z,W\rangle_t=\rho t$   with  $-1\leq \rho\leq 1.$

Define  $\hat{\mathbb{P}}$ 
as discussed in Section \ref{sec:LETF_dynamics},
then the process $\hat{B}_t$ given by
\begin{equation*} 
\hat{B}_t:=-\alpha\beta\sigma t+B_t
\end{equation*} 
is a $\mathbb{\hat{\mathbb{P}}}$-Brownian motion.
From equation \eqref{eqn:expected_utility_of_LETF_hat_P}, it follows that 
$$\mathbb{E}^\mathbb{P}[L_t^\alpha]=\mathbb{E}^\mathbb{\hat{\mathbb{P}}}[e^{-\alpha(\beta-1)\int_0^tr_s\,ds}]
\,e^{\alpha\beta\mu t-\frac{1}{2}\alpha(1-\alpha)\beta^2\sigma^2t}\;.$$
The $\hat{\mathbb{P}}$-dynamics of $r_t$ is
$$dr_t=(\theta+\alpha\beta\delta\sigma\rho-a r_t)\,dt+\delta \,d\hat{Z}_t$$
with   a $\hat{\mathbb{P}}$-Brownian $\hat{Z}_t.$

We now explore the martingale extraction of
$e^{-\alpha(\beta-1)\int_0^tr_s\,ds}$ with the process $r_t$ playing the role of $G_t$ in Section \ref{sec:martingale_extraction}.
Consider the infinitesimal generator $\mathcal{L}$ of the diffusion $r_t$ with killing rate  $\alpha(\beta-1)r_t.$
We know that the generator $\mathcal{L}$ is
$$\mathcal{L}\phi(r)=\frac{1}{2}\delta^2\phi''(r)
+(\theta+\alpha\beta\delta\sigma\rho-a r)\,\phi'(r)-\alpha(\beta-1)r\,\phi(r)\;.$$
It can be shown that 
\begin{equation*}
(\lambda,\phi(r)):=\left(\frac{1}{2a^2}\alpha(1-\beta)(-\alpha\delta^2(1-\beta)+2a(\theta+\alpha\beta\delta\sigma\rho)),e^{-\frac{\alpha(1-\beta)r}{a} }\right)
\end{equation*}
is an admissible eigenpair of $\mathcal{L}.$
The process $r_t$ satisfies
$$dr_t=(\theta+\alpha\beta\delta\sigma\rho- {\alpha\delta^2(1-\beta)/a} -a r_t)\,dt+\delta \,dW_t\;,$$
where $W_t$ is a Brownian motion under
the corresponding transformed measure $\mathbb{Q}.$
It follows that
\begin{align*}
\mathbb{E}^\mathbb{P}[L_t^\alpha]
&=\mathbb{E}^\mathbb{\hat{\mathbb{P}}}[e^{-\alpha(\beta-1)\int_0^tr_s\,ds}]
 \,e^{\alpha\beta\mu t-\frac{1}{2}\alpha(1-\alpha)\beta^2\sigma^2t}\\
&=\mathbb{E}^\mathbb{Q}[e^{\kappa r_t}]
 \,e^{(\alpha\beta\mu-\frac{1}{2}\alpha(1-\alpha)\beta^2\sigma^2+ \frac{1}{2a^2}\alpha^2\delta^2(1-\beta)^2-
\frac{1}{a}\alpha(1-\beta)(\theta+\alpha\beta\delta\sigma\rho))t-\kappa r_0}
\end{align*}
and we  know that
$\mathbb{E}^\mathbb{Q}[e^{\kappa r_t}]$
converges to a positive constant because $r_t$ is again an OU process under $\mathbb{Q}.$
In conclusion, we have the following proposition.
\begin{proposition} Suppose that the reference price  process $X_t$ and the interest rate $r_t$ satisfy   \eqref{eqn:Vasicek} and  \eqref{eqn:Vasicek2} respectively. Then, we have 	\begin{equation}\label{limitgbmv}
	\lim_{t\rightarrow\infty}\frac{1}{t}\log \mathbb{E}^\mathbb{P}[L_t^\alpha]
	=\alpha\beta\mu-\frac{1}{2}\alpha(1-\alpha)\beta^2\sigma^2+ \frac{1}{2a^2}\alpha^2\delta^2(1-\beta)^2-
	\frac{1}{a}\alpha(1-\beta)(\theta+\alpha\beta\delta\sigma\rho).
	\end{equation}  
 \end{proposition}

We now find the optimal leverage ratio $\beta^*$ for a static investor.
To examine the dependence of the limit \eqref{limitgbmv} on $\beta$, we define 
$$\Lambda(\beta)=C_1\beta^2+C_2\beta,$$
with the constants
$$C_1=-\frac{1}{2}\alpha(1-\alpha)\sigma^2+\frac{\alpha^2\delta^2}{2a^2}+\frac{\alpha^2\delta\sigma\rho}{a}\;,\;C_2=\alpha\mu
-\frac{\alpha^2\delta^2}{a^2}+\frac{\alpha\theta}{a}-\frac{\alpha^2\delta\sigma\rho}{a}\;.$$
Note that  $\Lambda(\beta)$ is a quadratic function.
If  $C_1<0,$ then $\beta^*=-\frac{C_2}{2C_1}$ is optimal.
If $C_1>0,$ then a more positive  (resp. more negative) $\beta$ is always more favorable  when 
$\frac{C_2}{2C_1}>0$ (resp.  $\frac{C_2}{2C_1}<0$).
In the special case with  $C_1=0,$ then a more positively    (resp. more negatively) leveraged ETF is preferred when 
$C_2>0$ (resp. $C_2<0$).

\begin{figure}\label{fig:Vas}\begin{center} \includegraphics[trim={0cm 0cm 0cm 0cm},clip,scale = 0.6]{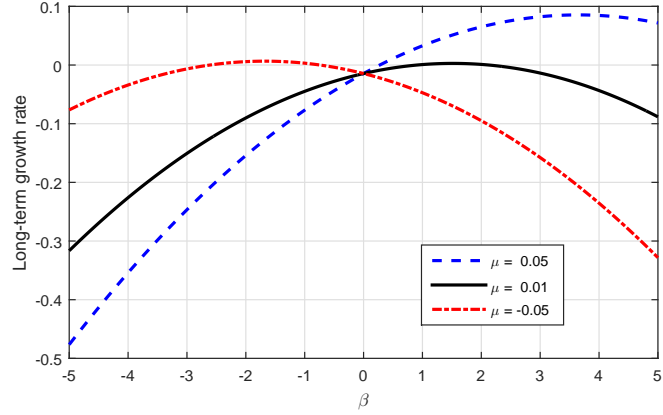} \caption{Long-term growth rates of expected utility under the GBM model with  Vasicek interest rate corresponding to three values of $\mu\in\{0.05, 0.01, -0.05\}.$ The optimal leverage ratios $\beta^*$ (maximizers of these curves)  are $\{3.65,  1.52, -1.68\},$  respectively. } \label{Fig_gbmv}\end{center} \end{figure}

In Figure  \ref{Fig_gbmv}, we display     the long-term growth rate \eqref{limitgbmv} as a function of $\beta$ for different values of $\mu$.   The parameters are: $\alpha=0.8,$   $r=0.01,$ $\theta=0.16,$ $\delta=0.89,$ $
a=3,$ $\sigma=0.3$, and $\rho=-0.5$. We can see that a positive (resp. negative) leverage ratio $\beta^*$ is optimal in a bull (resp. bear) market  with  $\mu=0.05$ (resp.  $\mu=-0.05$).  A positively leveraged ETF   with $\beta^* = 1.52$ is preferred here even when the reference asset offers no  excess return (i.e. $\mu -r = 0)$. This presents   an interesting contrast to  the Heston model depicted in Figure \ref{fig_hest} and the GBM model whereby the optimal leverage ratio $\beta^*=0$ whenever $\mu=r$. 

\subsection{Inverse GARCH interest rate}
\label{sec:r_I_GARCH}
Another model for the stochastic short interest rate is  
the inverse GARCH diffusion model, which was discussed in  Section \ref{sec:inv_GARCH}. Suppose that the reference price    $X_t$ and the short interest rate $r_t$ satisfy the SDEs 
\begin{equation}\label{eqn:inverse_GARCH}
\begin{aligned}
&dX_t=\mu X_t\, dt+ \sigma X_t\,dB_t\;,\\ 
&dr_t=(\theta-a r_t)r_t\,dt+\delta r_t\,dZ_t\;, 
\end{aligned} 
\end{equation}
where $B_t$ and $Z_t$ are two  Brownian motions such that  $\langle Z,W\rangle_t=\rho t$  with    $-1\leq \rho\leq 1.$ We assume that  $\mu,a,\delta>0$ and $\theta>\delta^2.$

Following the procedure in Section \ref{sec:LETF_dynamics}, we define the measure $\hat{\mathbb{P}}$, and express under this measure  the expected utility 
$$\mathbb{E}^\mathbb{P}[L_t^\alpha]=\mathbb{E}^\mathbb{\hat{\mathbb{P}}}[e^{-\alpha(\beta-1)\int_0^tr_s\,ds}]
\,e^{\alpha\beta\mu t-\frac{1}{2}\alpha(1-\alpha)\beta^2\sigma^2t}\;.$$
The stochastic interest rate evolves according to 
\begin{equation}\label{eqn:r_GAR}
dr_t=(\theta+\alpha\beta\delta\sigma\rho-a r_t)r_t\,dt+\delta r_t \,d\hat{Z}_t\;,
\end{equation}
where $\hat{Z}_t$ is a $\hat{\mathbb{P}}$-Brownian motion.

We now present  the martingale extraction of
$e^{-\alpha(\beta-1)\int_0^tr_s\,ds}.$
The   infinitesimal generator $\mathcal{L}$  of $r_t$ is
$$\mathcal{L}\phi(r)=\frac{1}{2}\delta^2r^2\phi''(r)
+(\theta+\alpha\beta\delta\sigma\rho-a r)r\,\phi'(r)-\alpha(\beta-1)r\,\phi(r)\;.$$
It can be verified  that 
\begin{equation*}
(\lambda,\phi(r))=\left(-\frac{1}{2a^2}\alpha\delta^2(\beta-1)(\alpha\beta-\alpha+a)+\frac{1}{a}\alpha(\beta-1)(\theta+\alpha\beta\delta\sigma\rho),r^{\alpha(1-\beta)/a}\right)
\end{equation*}
is an admissible eigenpair of $\mathcal{L}$
provided  $\theta+\alpha\beta\delta\sigma\rho-\alpha\delta^2(\beta-1)/a>\delta^2$, which   explains our condition on the parameters.

After a change to the  transformed measure $\mathbb{Q}$, $r_t$ satisfies
\begin{equation}\label{eqn:GARCH_r}
dr_t=(\theta+\alpha\beta\delta\sigma\rho-\alpha\delta^2(\beta-1)/a-a r_t)r_t\,dt+\delta r_t \,dW_t\;, 
\end{equation}
where $W_t$ is a Brownian motion under  $\mathbb{Q}.$
Connecting the measures through the expected utility, we write 
\begin{align}
 &\mathbb{E}^\mathbb{P}[L_t^\alpha]\\
=&\mathbb{E}^\mathbb{\hat{\mathbb{P}}}[e^{-\alpha(\beta-1)\int_0^tr_s\,ds}]
\,e^{\alpha\beta\mu t-\frac{1}{2}\alpha(1-\alpha)\beta^2\sigma^2t}\\
=&\mathbb{E}^\mathbb{Q}[r_t^{-\alpha(1-\beta)/a}]
\,e^{(\alpha\beta\mu -\frac{1}{2}\alpha(1-\alpha)\beta^2\sigma^2+ \frac{1}{2a^2}\alpha\delta^2(\beta-1)(\alpha\beta-\alpha+a)-\frac{1}{a}\alpha(\beta-1)(\theta+\alpha\beta\delta\sigma\rho))t}\,r_0^{\alpha(1-\beta)/a}\;.\label{eqn:I_G}
 \end{align}
Inspecting the last line \eqref{eqn:I_G}, we point out  that 
\begin{equation}\label{eqn:II_G}\left\{\,
\begin{aligned}
\lim_{t\rightarrow\infty}&\mathbb{E}^\mathbb{Q}[r_t^{-\alpha(1-\beta)/a}]=\textnormal{(positive const)}&&\textnormal{if }\, \alpha(1-\beta)/a+\frac{2}{\delta^2}(\theta+\alpha\beta\delta\sigma\rho)-1>0\,,\\
&\mathbb{E}^\mathbb{Q}[r_t^{-\alpha(1-\beta)/a}]=\infty &&\textnormal{otherwise}\;.
\end{aligned}\right.
\end{equation}
We refer  to Appendix \ref{app:inv_GARCH} for the    equality  in  \eqref{eqn:I_G} and two equalities in \eqref{eqn:II_G}.  

 \begin{proposition} Consider  the reference  price  $X_t$  and the short interest rate $r_t$ that satisfy  equation \eqref{eqn:inverse_GARCH}.	
If  $\alpha(1-\beta)/a+\frac{2}{\delta^2}(\theta+\alpha\beta\delta\sigma\rho)-1>0$, then 
\begin{equation} \label{limitiggbm}
\lim_{t\rightarrow\infty}\frac{1}{t}\log \mathbb{E}^\mathbb{P}[L_t^\alpha]
=\alpha\beta\mu-\frac{1}{2}\alpha(1-\alpha)\beta^2\sigma^2+ \frac{1}{2a^2}\alpha^2\delta^2(1-\beta)^2-
\frac{1}{a}\alpha(1-\beta)(\theta+\alpha\beta\delta\sigma\rho).
\end{equation} 
Otherwise, we have \[\lim_{t\rightarrow\infty}\frac{1}{t}\log \mathbb{E}^\mathbb{P}[L_t^\alpha]=\infty.\] 
\end{proposition}

Using this result, we can find  the optimal leverage ratio  $\beta^*$ for a long-term investor by analyzing 
the limit  \eqref{limitiggbm} as a function of $\beta$, namely, 
\[\Lambda(\beta):=\alpha\beta\mu-\frac{1}{2}\alpha(1-\alpha)\beta^2\sigma^2+ \frac{1}{2a^2}\alpha^2\delta^2(1-\beta)^2-
\frac{1}{a}\alpha(1-\beta)(\theta+\alpha\beta\delta\sigma\rho).\]
The function $\Lambda(\beta)$ is quadratic  provided  that
$\alpha(1-\beta)/a+\frac{2}{\delta^2}(\theta+\alpha\beta\delta\sigma\rho)-1>0$  
on $|\beta|\leq3$. The procedure to   determine the maximum of $\Lambda(\beta)$ is the same as that presented in     Section \ref{sec:Vasicek}, and is thus omitted.

\section{Quadratic models}
\label{sec:Q}
In this section, we consider  a quadratic model given by 
$X_t=e^{|Y_t|^2},$
where $Y_t$ is a $d$-dimensional Ornstein-Uhlenbeck (OU) process  
\begin{align}dY_t=(b+BY_t)\,dt+\sigma\,dB_t\;,\quad \;Y_0=0_d\;.\label{quadsde} \end{align}
Here, $b$ is a $d$-dimensional column vector, $B$ is a $d\times d$ matrix, and $\sigma$ is a non-singular $d\times d$ matrix, so that $a=\sigma\sigma^{\top}$ is strictly positive definite. We refer to \cite{ahn2002quadratic} and \cite{QinLinetsky2015} for more details about this quadratic model.
The   interest rate $r$  is assumed to be a positive  constant. Under the qadratic model, the  LETF price $L_t$ can be expressed  as  
$$L_t={X_t}^\beta e^{-r(\beta-1)t-2\beta(\beta-1)\int_0^tY_u^\top a Y_u\,du}\;,$$
which is derived in Appendix \ref{app:quadratic_models}.
From equation \eqref{eqn:expected_utility_of_LETF_hat_P}, the expected utility  is given by 
\begin{equation*}
\begin{aligned}
\mathbb{E}^\mathbb{P}[L_t^\alpha]
&=\mathbb{E}^\mathbb{P}
[e^{-2\alpha\beta(\beta-1)\int_0^tY_u^\top a Y_u\,du}\,X_t^{\alpha\beta}]\,e^{(r\alpha(1-\beta))t}\\
&=\mathbb{E}^\mathbb{P}
[e^{-2\alpha\beta(\beta-1)\int_0^tY_u^\top a Y_u\,du}\,e^{\alpha\beta|Y_t|^2}]\,e^{(r\alpha(1-\beta))t}\;. 
\end{aligned}
\end{equation*} 

We now apply our   martingale extraction method developed in Section \ref{sec:martingale_extraction}. The    infinitesimal generator $\mathcal{L}$ corresponding to $Y_t$ in \eqref{quadsde}  is
$$\mathcal{L}\phi(y)=\nabla \phi(y) (b+By)+\frac{1}{2}\sum_{i,j}(H\phi(y))_{ij}\,a_{ij}
-2\alpha\beta(\beta-1)y^\top a y\,\phi(y)\;,$$
where $\nabla \phi(y)$ is the gradient row vector and $H\phi(y)$ is the Hessian matrix.
We can find an admissible eigenpair of $\mathcal{L}$ by the following way.
Let $V$ be the {\em stabilizing solution} (i.e., $V$ is symmetric, $B-2aV$ is non-singular and the eigenvalues of $B-2aV$ have negative real parts) of 
\begin{equation}\label{eqn:stabilizing}
2VaV-B^\top V-VB-2\alpha\beta(\beta-1)a=0\;, 
\end{equation}  
and define a vector $u$ by
\begin{equation}\label{eqn:u_stabilizing}
u=2(2a-V^{-1}B^\top)^{-1}b\;. 
\end{equation}
In this case, we obtain an admissible eigenpair of $\mathcal{L}$, given by 
$$(\lambda,\phi(y))=(-\frac{1}{2}u^\top au+tr(aV)+u^\top b\,,\,e^{-u^\top y-y^\top V y}).$$
This leads to the  martingale extraction
of $$e^{-2\alpha\beta(\beta-1)\int_0^tY_u^\top a Y_u\,du}\,.$$ Refer to Section 6.2.2 in \cite{QinLinetsky2015} for more details about this eigenpair.
Denote by  $\mathbb{Q}$   the transformed measure   with respect to $(\lambda,\phi)$ (see \eqref{QA}).
Under the measure  $\mathbb{Q}$,   $X_t$  evolves according to 
\begin{equation}\label{eqn:SDE_QTSM_under_P}
dX_t=(b-au+(B-2aV)X_t)\,dt+\sigma\,dW_t\,,
\end{equation}
where $W_t$ is a Brownian motion under $\mathbb{Q}.$

The expected utility from holding the LETF can be expressed as 
\begin{align*}
\mathbb{E}^\mathbb{P}[L_t^\alpha]
&=\mathbb{E}^\mathbb{P}
[e^{-2\alpha\beta(\beta-1)\int_0^tY_u^\top a Y_u\,du}\,e^{\alpha\beta|Y_t|^2}]\,e^{(r\alpha(1-\beta))t}\\
&=\mathbb{E}^\mathbb{Q}
[e^{u^\top Y_t+Y_t^\top V Y_t+\alpha\beta|Y_t|^2}]\,e^{(r\alpha(1-\beta)+\frac{1}{2}u^\top au-tr(aV)-u^\top b)t}\;.
\end{align*}
For any fixed $t$,  $Y_t$ is a multivariate normal random variable. Therefore, we compute explicitly the expectation    
$$\mathbb{E}^\mathbb{Q}
[e^{u^\top Y_t+Y_t^\top V Y_t+\alpha\beta|Y_t|^2}]
=\frac{1}{\sqrt{(2\pi)^d\det\Sigma_t}}\int_{\mathbb{R}^d} e^{u^\top y+y^\top Vy+\alpha\beta|y|^2-\frac{1}{2}(y-\mu_T)^\top\Sigma_t^{-1}(y-\mu_t)}\,dy\;,$$ 
where $\mu_t$ and $\Sigma_t$ are the mean vector and the covariance matrix of $Y_t$ under $\mathbb{Q},$ respectively.
Under $\mathbb{Q},$ the coefficient of $Y_t$ in the drift term of equation \eqref{eqn:SDE_QTSM_under_P} is  $B-2aV,$ all of whose eigenvalues have negative real parts.
Thus, the distribution of $Y_t$ is convergent to an invariant distribution, which is a non-degenerate multivariate normal random variable.  
Let $\Sigma_\infty$ be the covariance matrix of the invariant distribution. 
Let $C:=V+\alpha\beta I_d-\frac{1}{2}\Sigma_\infty,$ which is a symmetric matrix.
The convergence and divergence of the integral on the right-hand side depends on the eigenvalues of $C.$
We summarize as follows:

\begin{equation*}\left\{\,
\begin{aligned}
\lim_{t\rightarrow\infty}&\mathbb{E}^\mathbb{Q}
[e^{u^\top Y_t+Y_t^\top V Y_t+\alpha\beta|Y_t|^2}]=\textnormal{(positive constant)}\;&&\textnormal{if all eigenvalues of }C\textnormal{ are negative}\,, \\
&\mathbb{E}^\mathbb{Q}
[e^{u^\top Y_t+Y_t^\top V Y_t+\alpha\beta|Y_t|^2}]=\infty &&\textnormal{otherwise}\,.
\end{aligned}\right.
\end{equation*}
\begin{proposition}Let $V$ be the stabilizing solution of equation \eqref{eqn:stabilizing},  and $u$ be defined by equation \eqref{eqn:u_stabilizing}. Then,   the long-term growth rate  is given by
 	\begin{equation*}
	\begin{aligned}
	&\lim_{t\rightarrow\infty}\frac{1}{t}\log \mathbb{E}^\mathbb{P}\left[L_t^\alpha\right] 
	\\
	=&\left\{\,
	\begin{aligned}
	r\alpha(1-\beta)+\frac{1}{2}u^\top au&-tr(aV)-u^\top b&&\textnormal{if all eigenvalues of }C\textnormal{ are negative}\,,\\
	&\infty &&\textnormal{otherwise}\,.\\
	\end{aligned}\right.
	\end{aligned}
	\end{equation*} 
\end{proposition}

\section{Conclusions}
\label{sec:conclusion}

In our study of  the long-term growth rate  of  expected utility of LETF, we propose the martingale extraction approach  and turn a path-dependent expectation into a path-independent one that is significantly more amenable for analysis and leads to explicit solutions. In determining the long-term growth rate of expected utility (or expected return), we  also illustrate and solve  the embedded eigenpair (eigenvalue and eigenfunction) problems. In each of the  single-factor and multi-factor    models studied herein, we derive the eigenpair  as well as the limit of the growth rate.  Using the formula for the  long-term growth rate, we also determine the optimal leverage ratio  and examine the effects of various model parameters. The results are  useful not only for individual or institutional investors, but also ETF providers and regulators as they ought to know the long-term performance for any LETF traded in the market.  

There are a number of directions for future research. One  direction is to investigate the long-term price behavior of options written on LETFs (see e.g. \cite{LeungSircar2015, LeungLorig2016}). Recent studies on the long-term price behavior of options can be found in \cite{Park2016}.  Given that the martingale extraction method studied herein works very well for  LETFs which are constant proportion portfolios, one interesting and practical  extension  is to  adapt it to other dynamic portfolio strategies.

\appendices

\section{The extended CIR model}
\label{app:the_mart_extrac_CIR}
We evaluate the limit stated in   \eqref{eqn:ext_CIR_lim} under  the extended CIR model. We recall the $\mathbb{Q}$-dynamics of $X_t$   in   \eqref{eqn:ext_CIR_Q}:
\begin{equation} 
dX_t=( \ell-\mu X_t)\,dt+\sigma\sqrt{X_t}\,dW_t\;, 
\end{equation}
with $\ell:=\theta+\kappa\sigma^2$.

\begin{proposition}
	For $p\in\mathbb{R},$ we have 
	\begin{equation*}
	\lim_{t\rightarrow\infty}\frac{1}{t}\log \mathbb{E}^\mathbb{Q}[X_t^{p}\,e^{\frac{2\mu}{\sigma^2}X_t}]=\left\{
	\begin{aligned}
	&\Big(p+\frac{2\ell}{\sigma^2}\Big)\mu\quad&&\textnormal{ if }\; p+\frac{2\ell}{\sigma^2}>0\;,\\
	&\quad\infty&&\textnormal{ if }\; p+\frac{2\ell}{\sigma^2}\leq0\;.\\
	\end{aligned}\right.
	\end{equation*} 
\end{proposition}
\noindent Before proving this proposition, we define the following notation.\newline

\noindent {\bf Notation.} Let $p(x)$ and $q(x)$ be two positive functions of $x.$
Denote this by $$p(x)\simeq q(x)\quad \textnormal{ at } x=x_0$$ if $\lim_{x\rightarrow x_0}{p(x)}/{q(x)}$ exists and is a nonzero constant. 
We denote this by
$$p\lesssim q$$ if there exists a positive constant $c$ such that $p(x)\leq c\cdot q(x).$ \newline

\begin{proof}
	The density function $g(x;t)$ of  $X_t$ at a fixed time $t$ is known to be
	\begin{equation}\label{eqn:CIR_density}
	g(x;t)=h_t\,e^{-u-v}\left(\frac{v}{u}\right)^{q/2}I_q(2\sqrt{uv})\;,
	\end{equation}
	where $I_q$ is the modified Bessel function of the first kind of order $q$, and 
	$$h_t=\frac{2\mu }{\sigma^2(1-e^{-\mu t})}\,,\;q=\frac{2 \ell}{\sigma^2}-1\,,\;
	u=h_te^{-\mu t}\,,\;v=h_tX_0\;.$$
	After rewriting slightly, we find
	$$g(x;t)=k_t\,h_t\,e^{-h_tx}x^{q/2}I_q(2h_te^{-\mu t/2}\sqrt{x})\;.$$
	Here, $k_t=e^{-h_te^{-\mu  t}}e^{\mu  q t/2}.$
	The quantity 
	$$\lim_{t\rightarrow\infty}\frac{1}{t}\log \int_0^\infty x^{p}e^{\frac{2\mu }{\sigma^2}x}\,g(x;t)\,dx$$
	is of interest to us.

By inspection, we obtain 
\begin{align}\label{appa1}\int_0^\infty\!\!x^{p+q}e^{-p_t x}\,dx
	\lesssim \int_0^\infty\!\! x^{p}e^{\frac{2\mu }{\sigma^2}x}\,g(x;t)\,dx
	\lesssim \int_0^\infty\!\! x^{p+q}e^{-p_tx}e^{2h_te^{-\mu  t/2}\sqrt{x}}\,dx\,,\end{align}
	where $p_t=h_t-\frac{2\mu }{\sigma^2}$.
	This follows  from $z^q\lesssim I_q(z)\lesssim z^{q}e^z.$
	We now show that  if $p+q+1>0,$ then 
	$$\textnormal{(right- and left-hand sides of \eqref{appa1})}\simeq e^{(p+q+1)a t}\;.$$
	For the right-hand side of \eqref{appa1}, substitute $y=p_tx,$ then 
	\begin{equation*}
	\begin{aligned}
	\int_0^\infty\!\! x^{p+q}e^{-p_tx}e^{2h_te^{-\mu  t/2}\sqrt{x}}\,dx
	=\,p_t^{-p-q-1}\int_0^\infty y^{p+q} e^{-y}e^{2h_t e^{-\mu t/2}p_t^{-1/2}\sqrt{y}}\,dy\;.
	\end{aligned}
	\end{equation*}
	As $t$ approaches to infinity, $h_t e^{-\mu t/2}p_t^{-1/2}$ converges to a constant, so  the integral term converges to a positive constant.
	By direct calculation, $p_t^{-p-q-1}\simeq e^{(p+q+1)\mu t}.$
	This implies that $$\textnormal{(right-hand side of \eqref{appa1})}\simeq e^{(p+q+1)\mu t}\;.$$
	The proof is similar for  the left-hand side of \eqref{appa1} given that  $p+q+1>0.$  	On the other hand, if $p+q+1\leq 0,$ then the left-hand side of \eqref{appa1} is infinity. 	This completes the proof.	
\end{proof}

\section{The inverse GARCH model}
\label{app:inv_GARCH}

With refernce to Section \ref{sec:r_I_GARCH},  recall the $\hat{\mathbb{P}}$-dynamics of the stochastic interest rate
\begin{equation} 
dr_t=(\theta+\alpha\beta\delta\sigma\rho-a r_t)r_t\,dt+\delta r_t \,d\hat{Z}_t\;.
\end{equation}
We  investigate  the long-term growth rate of
$$\mathbb{E}^\mathbb{\hat{\mathbb{P}}}[e^{-\alpha(\beta-1)\int_0^tr_s\,ds}].$$
This in turn yields the equalities  in \eqref{eqn:I_G} and   \eqref{eqn:II_G}.
For convenience, in this appendix, we put
$$\hat{\mathbb{P}}\to\mathbb{P}\,,\;\hat{Z}_t\to B_t \,,\;\theta+\alpha\beta\delta\sigma\rho\to\theta\,,\;\delta\to \sigma\,,\;\alpha(\beta-1)\to c\;.$$

With these new parameters, we investigate the long-term growth rate of the expectation 
$$\mathbb{E}^\mathbb{P}[e^{-c\int_0^tr_s\,ds}]$$
when the process $r_t$ follows 
the inverse GARCH diffusion model:
\begin{equation*}
	\begin{aligned}
		dr_t=(\theta-a r_t)r_t\,dt+\sigma r_t\,dB_t\,,\;r_0=1
	\end{aligned}
\end{equation*}
with $a,\sigma>0$ and $\theta>\sigma^2.$ 
\begin{proposition} Let $\kappa:=c/a$ and assume that $\theta>(\kappa+1)\sigma^2.$ 
The long-term growth rate of the above expectation is given by 
$$\lim_{t\rightarrow\infty}\frac{1}{t}\log \mathbb{E}^\mathbb{P}[e^{-c\int_0^tr_s\,ds}]=-\theta\kappa+\frac{1}{2}\sigma^2\kappa(\kappa+1)\;.$$
\end{proposition}

To prove this proposition, we will apply the martingale extraction to $e^{-c\int_0^tr_s\,ds}.$
The corresponding 
infinitesimal generator is 
$$\mathcal{L}\phi(x)=\frac{1}{2}\sigma^2x^2\,\phi''(x)+(\theta-ax)x\,\phi'(x)-cx\,\phi(x)\;.$$
With $\kappa=c/a,$ it follows that the  corresponding eigenpair is 
$$(\lambda,\phi(x)):=\Big(\theta\kappa-\frac{1}{2}\sigma^2\kappa(\kappa+1)\,,\,x^{-\kappa}\Big)$$
is an admissible eigenpair of $\mathcal{L}$
when $\theta>(\kappa+1)\sigma^2.$ 
Let $\mathbb{Q}$ be the   corresponding transformed measure.
The process $r_t$ follows
$$dr_t=(\theta -\kappa\sigma^2-ar_t)r_t\,dt+\sigma r_t\,dW_t\;,$$
where $W_t$ is a $\mathbb{Q}$-Brownian motion.
Through the martingale extraction with respect to this eigenpair, the expectation is expressed by
$$\mathbb{E}^\mathbb{P}[e^{-c\int_0^tr_s\,ds}]=\mathbb{E}^\mathbb{Q}[r_t^\kappa]\,e^{(-\theta\kappa+\frac{1}{2}\sigma^2\kappa(\kappa+1))t}\;.$$
We show that
$$\lim_{t\rightarrow\infty}\mathbb{E}^\mathbb{Q}[r_t^{\kappa}]=\textnormal{(positive constant)}\;.$$
Under $\mathbb{Q},$ the process $Y_t=\frac{2a}{\sigma^2}r_t$ converges to the Gamma random variable with parameter $\gamma=\frac{2\theta}{\sigma^2}-2\kappa-1,$ that is,
then density function $p(y;t)$ of $Y_t$ 
converges to $p(y;\infty):=\frac{1}{\Gamma(\gamma)}y^{\gamma-1}e^{-y}$ 
as $t\rightarrow\infty$ (see Theorem 2.5 in \cite{Zhao2009} and Section 6.5.4 in \cite{Linetsky2004}) By the same argument in equation \eqref{eqn:GARCH_finite}, we know
\begin{align*}
\mathbb{E}^\mathbb{P}[r_t^{\kappa}]
=\Big(\frac{\sigma^2}{2a}\Big)^{\kappa}\,\mathbb{E}^\mathbb{P}[Y_t^{\kappa}]
&=\Big(\frac{\sigma^2}{2a}\Big)^{\kappa}\int_0^\infty y^{\kappa}\,p(y;t)\,dy \\
&\rightarrow
\Big(\frac{\sigma^2}{2a}\Big)^{\kappa}\frac{1}{\Gamma(\gamma)}\int_0^\infty y^{\frac{2\theta}{\sigma^2}-\kappa-2}e^{-y}\,dy\;. 
\end{align*}  
when $\theta>(\kappa+1)\sigma^2.$ This gives  the desired result.

\section{Quadratic models}
\label{app:quadratic_models}
We now derive the SDE for $X_t$  by  using the fromula $X_t=e^{|Y_t|^2}$, where 
$$dY_t=(b+BY_t)\,dt+\sigma\,dW_t\;,\;X_0=0_d\;.$$
Define $f(y)=e^{|y|^2},$ then the gradient row  vector and Hessian matrix are, respectively, 
\begin{align}\nabla f(y)=2f(y)y^\top\,, \label{ffy}\qquad Hf(y)=2f(y)
\begin{pmatrix}
1+2y_1^2& 2y_1y_2&\cdots& 2y_1y_d\\
2y_2y_1&1+2y_2^2&\cdots&2y_2y_d\\
\vdots& \vdots&\ddots &\vdots\\
2y_dy_1&2y_dy_2&\cdots &1+2y_d^2
\end{pmatrix}\;. 
\end{align}
By Ito's formula, it follows that
\begin{align}\label{dxxt}dX_t=df(Y_t)=\left(\nabla f(Y_t)(b+BY_t)+\frac{1}{2}\sum_{i,j}(Hf(Y_t))_{ij}(\sigma\sigma^\top)_{ij}\right)dt+\nabla f(Y_t)\sigma\,dW_t\;.\end{align}
Applying \eqref{ffy} to  \eqref{dxxt}, we get
$$\frac{dX_t}{X_t}=\ldots\,dt+2Y_t^\top\sigma\,dW_t\;,$$ where we have abstracted the drift term since its expression is not needed here.  From this and \eqref{lxxt}, the LETF price can be expressed as  
$$L_t={X_t}^\beta e^{-r(\beta-1)t-2\beta(\beta-1)\int_0^t|\sigma^\top Y_u|^2\,du}.$$
 From this we observe the path dependence of $L_t$ on the $d$-dimensional OU process $Y$.

\begin{small}
\begin{spacing}{0.5}
\bibliographystyle{apa}
\bibliography{LETF2015}

\begin{thebibliography}{}

\bibitem[\protect\astroncite{Ahn et~al.}{2002}]{ahn2002quadratic}
Ahn, D.-H., Dittmar, R.~F., and Gallant, A.~R. (2002).
\newblock Quadratic term structure models: Theory and evidence.
\newblock {\em Review of Financial Studies}, 15(1):243--288.

\bibitem[\protect\astroncite{Ahn and Gao}{1999}]{ahn1999parametric}
Ahn, D.-H. and Gao, B. (1999).
\newblock A parametric nonlinear model of term structure dynamics.
\newblock {\em Review of Financial Studies}, 12(4):721--762.

\bibitem[\protect\astroncite{Akian et~al.}{1999}]{akian1999}
Akian, M., Sulem, A., and Taksar, M.~I. (1999).
\newblock Dynamic optimisation of a long term growth rate for a mixed portfolio
  with transaction costs: The logarithmic utility case.
\newblock Technical Report RR-3626, INRIA.

\bibitem[\protect\astroncite{Avellaneda and Zhang}{2010}]{AZLETF}
Avellaneda, M. and Zhang, S. (2010).
\newblock Path-dependence of leveraged {ETF} returns.
\newblock {\em SIAM Journal of Financial Mathematics}, 1:586--603.

\bibitem[\protect\astroncite{Borovicka
  et~al.}{2011}]{BorovickaHansenHendricksScheinkman2011}
Borovicka, J., Hansen, L., Hendricks, M., and Scheinkman, J. (2011).
\newblock Risk price dynamics.
\newblock {\em Journal of Financial Econometrics}, 9(1):3--65.

\bibitem[\protect\astroncite{Carr and Sun}{2007}]{carr2007new}
Carr, P. and Sun, J. (2007).
\newblock A new approach for option pricing under stochastic volatility.
\newblock {\em Review of Derivatives Research}, 10(2):87--150.

\bibitem[\protect\astroncite{Cheng and Madhavan}{2009}]{ChengLETF}
Cheng, M. and Madhavan, A. (2009).
\newblock The dynamics of leveraged and inverse exchange-traded funds.
\newblock {\em Journal of Investment Management}, 4.

\bibitem[\protect\astroncite{Christensen and
  Wittlinger}{2012}]{christensen2012}
Christensen, S. and Wittlinger, M. (2012).
\newblock Optimal relaxed portfolio strategies for growth rate maximization
  problems with transaction costs.
\newblock {\em arXiv preprint arXiv:1209.0305}.

\bibitem[\protect\astroncite{Cox et~al.}{1985}]{cox1985theory}
Cox, J.~C., Ingersoll~Jr, J.~E., and Ross, S.~A. (1985).
\newblock A theory of the term structure of interest rates.
\newblock {\em Econometrica: Journal of the Econometric Society}, pages
  385--407.

\bibitem[\protect\astroncite{Fleming and Sheu}{1999}]{Fleming1999}
Fleming, W. and Sheu, S. (1999).
\newblock Optimal long term growth rate of expected utility of wealth.
\newblock {\em Annals of Applied Probability}, 9(3):871--903.

\bibitem[\protect\astroncite{Guasoni and Mayerhofer}{2016}]{guasoni2015limits}
Guasoni, P. and Mayerhofer, E. (2016).
\newblock The limits of leverage.
\newblock {\em Available at SSRN 2446817}.

\bibitem[\protect\astroncite{Hansen}{2012}]{Hansen2012}
Hansen, L.~P. (2012).
\newblock Dynamic valuation decomposition within stochastic economies.
\newblock {\em Econometrica}, 80(3):911--967.

\bibitem[\protect\astroncite{Hansen and Scheinkman}{2009}]{Hansen2009}
Hansen, L.~P. and Scheinkman, J.~A. (2009).
\newblock Long-term risk: An operator approach.
\newblock {\em Econometrica}, 77(1):177--234.

\bibitem[\protect\astroncite{Hata and Sekine}{2006}]{hata2006solving}
Hata, H. and Sekine, J. (2006).
\newblock Solving long term optimal investment problems with
  {C}ox-{I}ngersoll-{R}oss interest rates.
\newblock {\em Advances in Mathematical Economics}, pages 231--255.

\bibitem[\protect\astroncite{Heston}{1993}]{heston1993closed}
Heston, S.~L. (1993).
\newblock A closed-form solution for options with stochastic volatility with
  applications to bond and currency options.
\newblock {\em Review of Financial Studies}, 6(2):327--343.

\bibitem[\protect\astroncite{Hurd and Kuznetsov}{2008}]{hurd2008explicit}
Hurd, T. and Kuznetsov, A. (2008).
\newblock Explicit formulas for {L}aplace transforms of stochastic integrals.
\newblock {\em Markov Processes and Related Fields}, 14(2):277--290.

\bibitem[\protect\astroncite{Leung et~al.}{2016}]{LeungLorig2016}
Leung, T., Lorig, M., and Pascucci, A. (2016).
\newblock Leveraged {ETF} implied volatilities from {ETF} dynamics.
\newblock {\em Mathematical Finance}.
\newblock To appear.

\bibitem[\protect\astroncite{Leung and Santoli}{2012}]{LeungSantoli}
Leung, T. and Santoli, M. (2012).
\newblock Leveraged exchange-traded funds: Admissible leverage and risk
  horizon.
\newblock {\em Journal of Investment Strategies}, 2(1):39--61.

\bibitem[\protect\astroncite{Leung and Santoli}{2016}]{Leung2016}
Leung, T. and Santoli, M. (2016).
\newblock {\em Leveraged Exchange-Traded Funds: Price Dynamics and Options
  Valuation}.
\newblock SpringerBriefs in Quantitative Finance, Springer.

\bibitem[\protect\astroncite{Leung and Sircar}{2015}]{LeungSircar2015}
Leung, T. and Sircar, R. (2015).
\newblock Implied volatility of leveraged {ETF} options.
\newblock {\em Applied Mathematical Finance}, 22(2):162--188.

\bibitem[\protect\astroncite{Leung and Ward}{2015}]{LeungWard}
Leung, T. and Ward, B. (2015).
\newblock The golden target: Analyzing the tracking performance of leveraged
  gold {ETF}s.
\newblock {\em Studies in Economics and Finance}, 32(3):278--297.

\bibitem[\protect\astroncite{Lewis}{2000}]{Lewis2000}
Lewis, A. (2000).
\newblock {\em Option valuation under stochastic volatility: With {M}athematica
  code}.
\newblock Finance Press, Newport Beach, California.

\bibitem[\protect\astroncite{Linetsky}{2004}]{Linetsky2004}
Linetsky, V. (2004).
\newblock The spectral decomposition of the option value.
\newblock {\em International Journal of Theoretical and Applied Finance},
  7(3):337--384.

\bibitem[\protect\astroncite{Park}{2016}]{Park2016}
Park, H. (2016).
\newblock Sensitivity analysis of long-term cash flows.
\newblock {\em ArXiv preprint arXiv:1511.03744}.

\bibitem[\protect\astroncite{Pham}{2003}]{Pham2003}
Pham, H. (2003).
\newblock A large deviations approach to optimal long term investment.
\newblock {\em Finance and Stochastics}, 7(2):169--195.

\bibitem[\protect\astroncite{Pham}{2015}]{pham2015}
Pham, H. (2015).
\newblock Long time asymptotics for optimal investment.
\newblock In {\em Large Deviations and Asymptotic Methods in Finance}, pages
  507--528. Springer.

\bibitem[\protect\astroncite{Pinsky}{1995}]{Pinsky1995}
Pinsky, R. (1995).
\newblock {\em Positive harmonic functions and diffusion}, volume~45.
\newblock Cambridge University Press.

\bibitem[\protect\astroncite{Qin and Linetsky}{2015}]{QinLinetsky2015}
Qin, L. and Linetsky, V. (2015).
\newblock Positive eigenfunctions of {M}arkovian pricing operators:
  {H}ansen-{S}cheinkman factorization and {R}oss recovery.
\newblock {\em ArXiv preprint arXiv:1411.3075}.

\bibitem[\protect\astroncite{Tuckwell}{1974}]{tuckwell1974study}
Tuckwell, H. (1974).
\newblock A study of some diffusion models of population growth.
\newblock {\em Theoretical Population Biology}, 5(3):345--357.

\bibitem[\protect\astroncite{Vasicek}{1977}]{vasicek1977equilibrium}
Vasicek, O. (1977).
\newblock An equilibrium characterization of the term structure.
\newblock {\em Journal of Financial Economics}, 5(2):177--188.

\bibitem[\protect\astroncite{Zhao}{2009}]{Zhao2009}
Zhao, B. (2009).
\newblock Inhomogeneous geometric {B}rownian motion.
\newblock {\em Available at SSRN 1429449}.

\bibitem[\protect\astroncite{Zhu}{2014}]{Zhu2014}
Zhu, L. (2014).
\newblock Optimal strategies for a long-term static investor.
\newblock {\em Stochastic Models}, 30(3):300--318.

\end{thebibliography}
\end{spacing}
\end{small}

\end{document}